\documentclass[3p,12pt]{elsarticle}

\usepackage{lineno,hyperref}

\hypersetup{pdfauthor=author}

\modulolinenumbers[5]

\journal{Artificial Intelligence in Medicine}

\usepackage{subcaption}
\usepackage{soulutf8}
\usepackage[table]{xcolor}
\usepackage[normalem]{ulem}
\usepackage{booktabs}
\usepackage{numprint}
\usepackage[symbol]{footmisc}

\soulregister\cite7
\soulregister\ref7

\colorlet{green_bg}{green!25}
\colorlet{yellow_bg}{yellow!30}

\DeclareRobustCommand{\removed}{\bgroup\markoverwith{\textcolor{red}{\rule[0.5ex]{2pt}{1.5pt}}}\ULon}
\newcommand{\tabitem}{~~~-~}

\begin{document}


\begin{frontmatter}

\title{Simultaneous segmentation and classification of the retinal arteries and veins from color fundus images\footnote[2]{Accepted by the journal Artificial Intelligence in Medicine. DOI: \url{https://doi.org/10.1016/j.artmed.2021.102116}}}


\author[citic,varpa_inibic]{José Morano\corref{correspondingauthor}}
\cortext[correspondingauthor]{Corresponding author}
\ead{j.morano@udc.es}

\author[citic,varpa_inibic]{Álvaro S. Hervella}
\ead{a.suarezh@udc.es}

\author[citic,varpa_inibic]{Jorge Novo}
\ead{jnovo@udc.es}

\author[citic,varpa_inibic]{José Rouco}
\ead{jrouco@udc.es}

\address[citic]{Centro de Investigación CITIC, Universidade da Coruña, A Coruña, Spain}
\address[varpa_inibic]{VARPA Research Group, Instituto de Investigación Biomédica de A Coruña (INIBIC), Universidade da Coruña, A Coruña, Spain}

\begin{abstract}
    \paragraph{Background and objectives} The study of the retinal vasculature represents a fundamental stage in the screening and diagnosis of many high-incidence diseases, both systemic and ophthalmic. A complete retinal vascular analysis requires the segmentation of the vascular tree along with the classification of the blood vessels into arteries and veins. Early automatic methods approach these complementary segmentation and classification tasks in two sequential stages. However, currently, these two tasks are approached as a joint semantic segmentation, because the classification results highly depend on the effectiveness of the vessel segmentation.
    In that regard, we propose a novel approach for the simultaneous segmentation and classification of the retinal arteries and veins from eye fundus images.
    
    \paragraph{Methods} We propose a novel method that, unlike previous approaches, and thanks to the proposal of a novel loss, decomposes the joint task into three segmentation problems targeting arteries, veins and the whole vascular tree. This configuration allows to handle vessel crossings intuitively and directly provides accurate segmentation masks of the different target vascular trees.
    
    \paragraph{Results} The provided ablation study on the public Retinal Images vessel Tree Extraction (RITE) dataset demonstrates that the proposed method provides a satisfactory performance, particularly in the segmentation of the different structures. Furthermore, the comparison with the state of the art shows that our method achieves highly competitive results in the artery/vein classification, while significantly improving the vascular segmentation. 
    
    \paragraph{Conclusions} The proposed multi-segmentation method allows to detect more vessels and better segment the different structures, while achieving a competitive classification performance. Also, in these terms, our approach outperforms the approaches of various reference works.
    Moreover, in contrast with previous approaches, the proposed method allows to directly detect the vessel crossings, as well as preserving the continuity of both arteries and veins at these complex locations.

\end{abstract}


\end{frontmatter}



\section{Introduction}
\label{sec:introduction}

The analysis of the anatomical structures of the retina represents a key step in the diagnosis and screening of many relevant diseases \cite{Kanski:Ophthalmology:Elsevier:2011, Abramoff:Retinal:RBE:2010}. This is due to the significant changes that some of these structures manifest when a certain disease is present. Thus, for example, ophthalmic and systemic diseases like Age-Related Macular Degeneration (AMD), glaucoma, and diabetes provoke significant alterations in the retinal vessels, the optic disc or the macula. The abnormal growth of new blood vessels within or under the macula is representative of the AMD disease\cite{Rosenfeld:AMD:2006}. Microaneurysms, retinal haemorrhages, exudates and neovascularization are representative of Diabetic Retinopathy (DR) \cite{Kanski:Ophthalmology:Elsevier:2011}, as well as particular changes in the diameters of small retinal vessels can be a sign of diabetes \cite{daSilva:IJGM:2015}. The increased venous diameter and venous oxygen saturation is associated with glaucomatous damage \cite{Vandewalle:Glaucoma:AO:2014,Ramm:Vessels:GACEO:2014}. The generalized retinal arteriolar narrowing is significantly associated with the optic nerve damage caused by open-angle glaucoma \cite{Wang:Diameter:CEO:2007, Mitchell:Diameters:O:2005}. Many of these disorders are of a great severity, and they can lead to the partial or total loss of vision \cite{Abramoff:Retinal:RBE:2010, Mathers:Diseases:2006, Ross:Atherosclerosis:1999, Prokofyeva:Epidemiology:2012}. Nevertheless, an early diagnosis and treatment can avoid or mitigate some of the most harmful consequences. In this regard, color retinography, an affordable and widely available eye fundus imaging technique, is commonly used for early diagnosis and pathological screening. Thereby, performing an accurate analysis of this type of images is crucial.

One of the most commonly analyzed structures of the retina is the vascular tree, since it provides particularly meaningful information for numerous diseases \cite{Kanski:Ophthalmology:Elsevier:2011, Rosenfeld:AMD:2006, Staal:DRIVE:2004, Patton:Retinal:JAnat:2005, Fraz:Segmentation:CMPB:2012}. Thus, a complete analysis of the retinal vascular tree requires the segmentation of the retinal vessels from eye fundus images as well as its classification into arteries and veins. The retinal vasculature segmentation allows the measurement of different features of the blood vessels (such as width, length and tortuosity) with proven relevance in evaluating and monitoring some of the aforementioned diseases (e.g. DR). Complementarily, vasculature classification allows measuring some arteries/veins specific features that supplement the previous data, such as the arteriolar and venular diameters \cite{Kawasaki:Hypertension:JH:2009}. Among other utilities, the measurement of these diameters allows the calculation of the Arteriolar-to-Venular diameter Ratio (AVR), that has been found useful for the diagnosis of multiple diseases, such as hypertension or diabetes \cite{Ikram:AVR:IOVS:2004, Sun:AVR:SO:2009, Hatanaka:Narrowing:EMB:2005}. In clinical practice, however, the application of these analyses is limited due to two main reasons: first, they are challenging, repetitive and time-consuming for the clinical specialists, and second, the results that are obtained by distinct experts commonly present significant differences, especially for the small vessels. In this context, automatic vasculature segmentation and classification methods emerged as an auspicious alternative.

Regarding the segmentation of the retinal vasculature, the early works were mainly based on either \textit{ad hoc} image processing techniques \cite{Staal:DRIVE:2004,Jiang:Adaptive:TPAMI:2003,Nain:Shape:MICCAI:2004,Tolias:Fuzzy:TMI:1998} or the use of hand-engineered image features along with traditional supervised learning methods, such as Artificial Neural Networks \cite{Sinthanayothin:Localisation:BJO:1999, Marin:SupervisedVS:TMI:2011}. Later, advances in deep learning motivated vasculature segmentation methods based on the patch-wise application of Convolutional Neural Networks (CNN) \cite{Liskowski:VS:TMI:2016}.
Currently, the state of the art in vascular segmentation is dominated by Fully Convolutional Neural Networks (FCNN), of varying architectures and training approaches, which allow to directly obtain a vascular segmentation map from a full size image \cite{Fu:DeepVessel:MICCAI:2016,Fu:VSFully:ISBI:2016,Dasgupta:VesselSegmentationFCNN:ISBI:2017,Jiang:FullyConvVS:CMIG:2018,Oliveira:FullyConvVS:ESA:2018,Jin:DUNet:KBS:2019}.

On the other hand, the classification of the retinal vessels into arteries and veins (A/V) followed a similar state of the art progression to vasculature segmentation, since the latter has always been conceived as a prior step to A/V classification. Methodologically, A/V classification works have evolved from \textit{ad hoc} and traditional supervised learning methods \cite{Relan:Retinal:EMBC:2013,Relan:EMBS:2014,Dashtbozorg:TIP:2014,Estrada:TMI:2015} to CNN-based methods \cite{Welikala:CBM:2017}. However, the overall approach of these works consisted on sequentially performing the vasculature segmentation followed by the A/V classification. Despite its wide adoption, the main drawback of this overall approach is that the errors in the vasculature segmentation are propagated to the classification stage. To avoid this issue, several recent works proposed to simultaneously address both tasks as a multiclass semantic segmentation problem with FCNNs \cite{Xu:A/V:BOE:2018, Girard:Joint:AIM:2019, Hemelings:A/V:CMIG:2019, Galdran:Uncertainty:ISBI:2019, Ma:Multi-task:MICCAI:2019, Kang:A/V:CMPB:2020}. 
Some of these works formulate the problem with three classes---background, artery and vein---~\cite{Xu:A/V:BOE:2018,Girard:Joint:AIM:2019,Ma:Multi-task:MICCAI:2019,Kang:A/V:CMPB:2020}. Others, instead, add a fourth class ``uncertain''~\cite{Hemelings:A/V:CMIG:2019,Galdran:Uncertainty:ISBI:2019}. This class, as specified in several annotated datasets~\cite{Hemelings:A/V:CMIG:2019,Hu:RITE:2013,Qureshi:ISCBMS:2013,Orlando:MICCAI:2018}, comprises all the vessels which experts cannot determine whether they are arteries or veins. The fourth class presented in~\cite{Hemelings:A/V:CMIG:2019} agrees with this definition. However, in~\cite{Galdran:Uncertainty:ISBI:2019}, the vessel crossings are also included in the uncertain class.
The crossings class comprises all the areas where a vein overlaps an artery or vice versa~\cite{Rothaus:IVC:2009}, and it is also commonly identified in A/V datasets~\cite{Hu:RITE:2013,Qureshi:ISCBMS:2013,Orlando:MICCAI:2018}. Along the article, in line with our approach, we will use the same nomenclature as
the one originally specified in the datasets, which does not include the vessel crossings in the uncertain class.
In that regard, previous alternatives \cite{Hemelings:A/V:CMIG:2019,Galdran:Uncertainty:ISBI:2019} handle vessels crossings and uncertain vessels in a counter-intuitive way, and give raise to incomplete segmentation maps for both arteries and veins, as the network is forced to decide to classify a vessel as one of the aforementioned classes.

In this work, we propose a novel approach for the simultaneous segmentation and classification of the retinal arteries and veins from eye fundus images using FCNNs. This approach decomposes these tasks into three related segmentation problems: the segmentation of arteries, veins, and vessels. To train the networks using this approach, we propose a novel loss named ``Binary Cross-Entropy by 3'' (BCE3) that combines three independent segmentation losses, one for each class of interest: arteries, veins and vessels. This approach manages uncertain vessels and vessel crossings in an intuitive way. On the one hand, uncertain vessels are discarded for the computation of the arteries and veins segmentation losses, but included for the computation of the vessels segmentation loss. In this way, the network can freely assign uncertain vessels to any of the artery and vein classes, while still receiving feedback for uncertain vessels in the vessels segmentation subtask. On the other hand, vessel crossings are included for the computation of the three independent losses. Thus, the networks are able to effectuate complete segmentation maps for the three classes: arteries, veins, and vessels. Moreover, the proposed setting allows the networks to detect the crossings between the vessels of different type, despite not being specifically trained for it, through the intersection (product) of the predicted arteries and veins segmentation masks.
To evaluate the potential and advantages of this approach, we performed an exhaustive comparative study with the current state-of-the-art approaches, which commonly use multi-class Cross-Entropy loss, for the A/V segmentation and classification tasks in a multi-class semantic segmentation setting. For this comparison, we employ the publicly available Retinal Images vessel Tree Extraction (RITE) dataset, which is considered the reference standard for the classification of retinal arteries and veins.


\subsection{State of the art}
\label{subsec:SOTA}

In the literature, several works have approached the segmentation of the retinal vasculature and, to a lesser extent, the classification of vessels into arteries and veins. Additionally, only in recent times, some works have approached both tasks simultaneously.

Regarding the vasculature segmentation, the initially proposed methods were chiefly based on \textit{ad hoc} image processing techniques. Adaptive local thresholding \cite{Jiang:Adaptive:TPAMI:2003}, deformable models \cite{Nain:Shape:MICCAI:2004}, vessel tracking \cite{Tolias:Fuzzy:TMI:1998} and ridge detection \cite{Staal:DRIVE:2004} are some representative examples. Moreover, traditional supervised learning methods, such as Artificial Neural Networks (ANNs) \cite{Sinthanayothin:Localisation:BJO:1999, Marin:SupervisedVS:TMI:2011} or k-nearest neighbors \cite{Staal:DRIVE:2004}, were also applied. Later, the use of deep learning, and specifically Convolutional Neural Networks (CNNs), meant an important advance \cite{Liskowski:VS:TMI:2016, Fu:DeepVessel:MICCAI:2016}. 
These networks allow to automatically learn features during the training, contrary to classical approaches, where the feature extraction had to be manually designed. Thus, these networks can be trained end-to-end, from the raw data to the target decisions, and commonly produce better results \cite{LeCun:Gradient:IEEE:1998, Krizhevsky:AlexNet:NIPS:2012}. Additionally, the use of Fully Convolutional Neural Networks (FCNNs) \cite{Long:FullyConv:CVPR:2015,Ronneberger:U-Net:MICCAI:2015}, that are only composed of local operators, allowed to directly handle image inputs of arbitrary sizes, without needing to iterate over them using patches. Currently, the state-of-the-art approaches on retinal vasculature segmentation utilize FCNNs \cite{Fu:VSFully:ISBI:2016, Dasgupta:VesselSegmentationFCNN:ISBI:2017, Jiang:FullyConvVS:CMIG:2018, Oliveira:FullyConvVS:ESA:2018, Jin:DUNet:KBS:2019}.

The automatic A/V classification has been less explored than the vasculature segmentation.
Also, until recently, it was conceived as a two-stage process, in which the vessel classification followed the vasculature segmentation.
In this way, the initial A/V classification works either assumed the presence of vasculature segmentation masks, or some vessel pixels that were manually annotated, \cite{Relan:Retinal:EMBC:2013,Relan:EMBS:2014,Dashtbozorg:TIP:2014,Zamperini:Features:CBMS:2012} or proposed their own automatic vasculature segmentation method as an independent first stage \cite{Estrada:TMI:2015,Welikala:CBM:2017}. 
Hence, the classification stage was only focused on classifying the pixels previously segmented as vessels.
Furthermore, early works typically restricted A/V classification to certain regions of interest, normally the region around the optic disc (OD) \cite{Zamperini:Features:CBMS:2012,Relan:Retinal:EMBC:2013,Relan:EMBS:2014}.

Zamperini et al. \cite{Zamperini:Features:CBMS:2012} employed supervised classifiers with color, contrast and position features for classifying a reduced set of central vessel points. This points were previously selected by experts and were located near the OD. Saez et al. \cite{Saez:CMPB:2012} proposed an unsupervised method based on k-means clustering for the classification of previously segmented vessels around the OD. Relan et al. \cite{Relan:Retinal:EMBC:2013} proposed a semi-automatic unsupervised method based on Gaussian Mixture Model-Expectation Maximization (GMM-EM) clustering, instead of k-means, due to its less dependency on the initialization. This method makes use of color features and, similarly to \cite{Zamperini:Features:CBMS:2012}, it needs a set of manually marked vessel points to work. Later, the same authors \cite{Relan:EMBS:2014} also proposed a supervised method based on Least Square-Support Vector Machine classifier. Their approach assumed the presence of previously extracted vascular masks near the OD. Dashtbozorg et al. \cite{Dashtbozorg:TIP:2014} introduced a classification method that follows a graph-based approach. They were one of the first to propose a method for classifying the whole vascular tree. However, their method used previously segmented vessel masks for extracting the vascular graph. Estrada et al., in \cite{Estrada:TMI:2015}, presented a semi-automatic graph-based framework that incorporates a global likelihood model for A/V classification. Like \cite{Dashtbozorg:TIP:2014}, their method classifies the whole vascular tree, but it implies the participation of experts to manually correct the errors of the primarily extracted graph. Afterward, Welikala et al. \cite{Welikala:CBM:2017} were the first to propose the use of a CNN for the classification step. They employed a patch-wise approach with patches centered on the centerline vessel pixels. For the extraction of these pixels, the authors utilized an unsupervised vessel segmentation approach based on a multi-scale line detector.

The two-stage approach followed by early works, however, presents an important drawback. Since the classification is done only for the pixels previously segmented as vessels, the final results heavily depend on the efficacy of the segmentation method, and the errors in it are propagated to the classification step.
In recent times, to avoid this problem, several works have approached the simultaneous segmentation and classification of the retinal vessels as a semantic pixel-level classification task \cite{Xu:A/V:BOE:2018, Girard:Joint:AIM:2019, Hemelings:A/V:CMIG:2019, Galdran:Uncertainty:ISBI:2019, Ma:Multi-task:MICCAI:2019, Kang:A/V:CMPB:2020}. To that end, all these works employ FCNNs, as they currently represent the state of the art for semantic segmentation \cite{Hao:Neurocomputing:2020}. Furthermore, all of them use some type of image preprocessing to facilitate the training of the networks. Beyond this common characteristics, the works differ in many points. One of the most relevant is the formulation of the problem. Xu et al. \cite{Xu:A/V:BOE:2018}, Girard et al. \cite{Girard:Joint:AIM:2019}, Ma et al. \cite{Ma:Multi-task:MICCAI:2019} and Kang et al. \cite{Kang:A/V:CMPB:2020} formulated the problem as a three-class semantic segmentation task---background, artery, vein---, while Hemelings et al. \cite{Hemelings:A/V:CMIG:2019} and Galdran et al. \cite{Galdran:Uncertainty:ISBI:2019} added a fourth class ``uncertain''.
In \cite{Hemelings:A/V:CMIG:2019}, this class only comprises the vessels of uncertain class, coinciding with the definition provided in several A/V segmentation datasets~\cite{Hemelings:A/V:CMIG:2019,Hu:RITE:2013,Qureshi:ISCBMS:2013,Orlando:MICCAI:2018}. However, in \cite{Galdran:Uncertainty:ISBI:2019}, the ``uncertain'' class also includes vessel crossings (areas where a vein overlaps an artery or vice versa).
In~\cite{Hemelings:A/V:CMIG:2019}, the crossing regions are labelled with the class of the upper vessel, so they are not handled in a special way. From these works \cite{Hemelings:A/V:CMIG:2019,Galdran:Uncertainty:ISBI:2019}, the approach proposed by Galdran et. al \cite{Galdran:Uncertainty:ISBI:2019} is the only one that does not ignore the pixels from the ``uncertain'' class in the computation of the loss.

Another important point in related works is the strategy that was used by each of them to alleviate the scarcity of annotated data. This problem, despite not being exclusive of medical imaging, is particularly notable in this field, since data annotation is usually a challenging process that must be performed by clinical experts. Beyond the typical data augmentation operations (i.e. affine and elastic transformations and color and intensity variations), more or less common to all of them, the works implement two different types of mechanisms: random patch extraction and transfer learning. Girad et al. \cite{Girard:Joint:AIM:2019} train the network using small patches randomly extracted from areas close to the vessels. Hemelings et al. \cite{Hemelings:A/V:CMIG:2019} train the network using large patches randomly extracted. Kang et al. \cite{Kang:A/V:CMPB:2020} use an encoder petrained on ImageNet based on the GoogLeNet deep network architecture \cite{Szegedy:CVPR:2015}. Ma et al. \cite{Ma:Multi-task:MICCAI:2019} use small random patches and an encoder pretrained on ImageNet and based on the ResNet CNN architecture \cite{He:CVPR:2016}. Finally, some of these works also propose specific mechanisms to mitigate the manifest misclassification of certain sections of vessels, e.g. the classification of a segment halfway down a vein as ``artery''. Girard et al. \cite{Girard:Joint:AIM:2019} propose a postprocessing method based on graph propagation, Ma et al. \cite{Ma:Multi-task:MICCAI:2019} include an spatial activation mechanism, and Kang et al. \cite{Kang:A/V:CMPB:2020} incorporate a category-attention weighted fusion (CWF) module and a graph based vascular structure reconstruction algorithm for postprocessing the result images.

The rest of the manuscript is organized as follows. In Section~\ref{sec:materials_and_methods}, we present the methodology for the A/V segmentation and classification task, including the description of the different approaches to compare, the network architecture, the data, and the training details. Then, in Section~\ref{sec:results_discussion}, we present a comprehensive evaluation and discussion of the presented approaches. Also, we expose the comparisons with the state-of-the-art works for both the vasculature segmentation and the A/V classification. Finally, in Section~\ref{sec:conclusions}, we present the main conclusions derived from the results and the potential future work.


\section{Materials and methods}
\label{sec:materials_and_methods}

The simultaneous segmentation and classification of the retinal arteries and veins (SSCAV) requires both to identify the blood vessels in the image and to differentiate them between arteries and veins. In this work, we approach this integrated vascular analysis using FCNNs that are trained to provide pixel-wise image segmentations. The proposed methods decomposes the SSCAV into the three semantic segmentations targeting arteries, veins, and vessels, as described in Section \ref{subsec:MS_approach}.
In order to evaluate the performance of this approach and quantify and demonstrate its advantages, we compare it with the traditional semantic segmentation approach followed by the state-of-the-art works. 
The details of this approach are explained in Section \ref{subsec:traditional_approach}.

\begin{figure}
    \centering
    \subfloat[]{\includegraphics[width=0.19\textwidth]{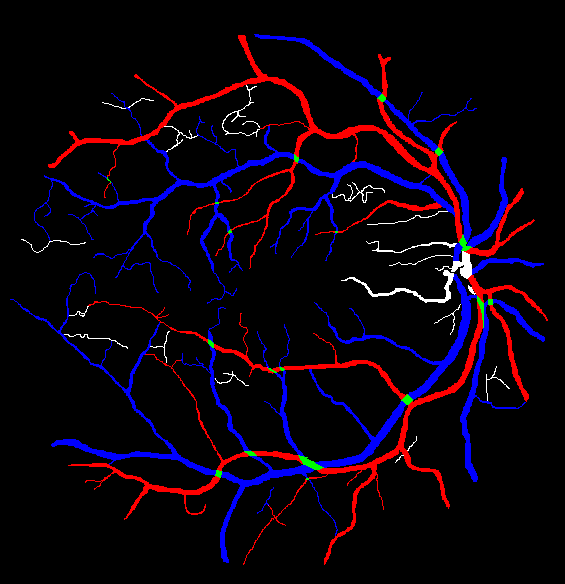}}
    \hfill
    \subfloat[]{\includegraphics[width=0.19\textwidth]{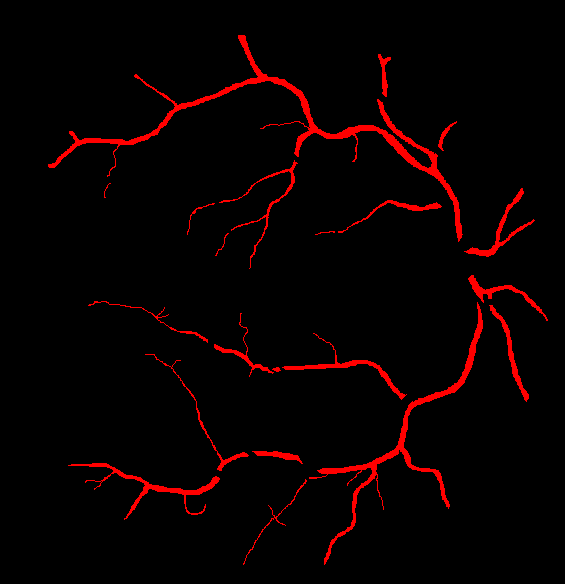}}
    \hfill
    \subfloat[]{\includegraphics[width=0.19\textwidth]{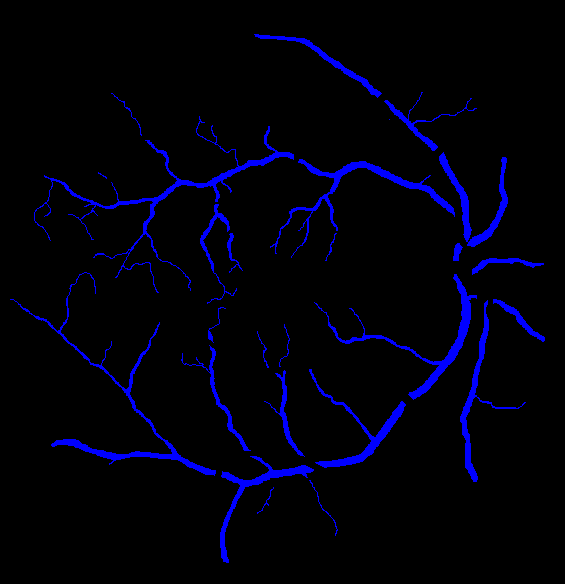}}
    \hfill
    \subfloat[]{\includegraphics[width=0.19\textwidth]{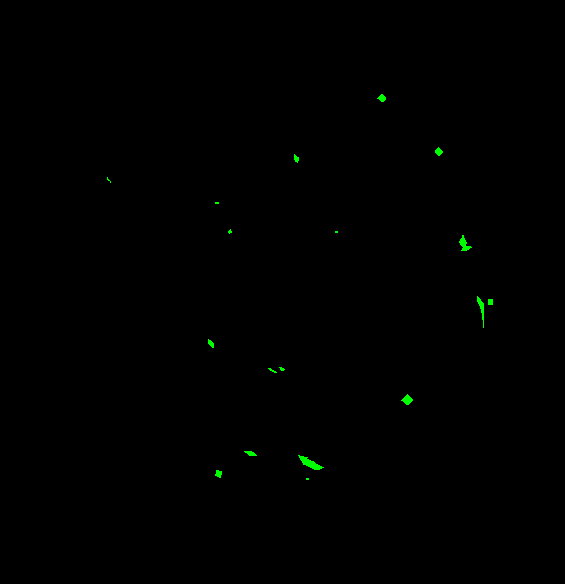}}
    \hfill
    \subfloat[]{\includegraphics[width=0.19\textwidth]{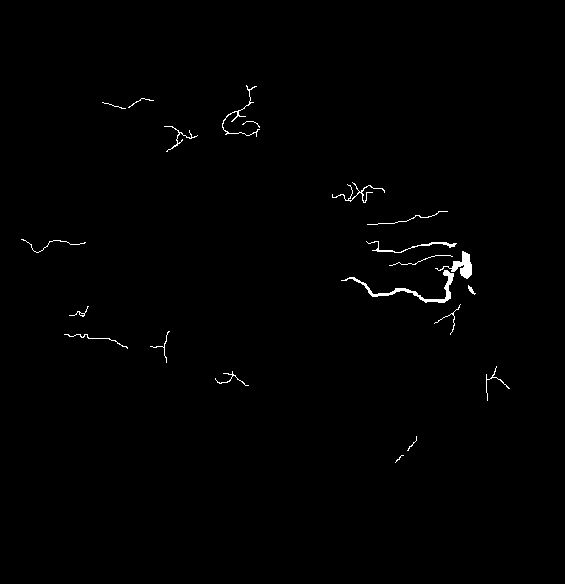}}
    \caption{Example RITE ground truth image (a) and its decomposition into (b)~arteries (red), (c)~veins (blue), (d)~crossings (green) and (e)~uncertain (white) classes.}
    \label{fig:example_img_dataset}
\end{figure}

The visual differentiation between arteries and veins can be difficult in some cases, like small vessels with poor contrast and not clearly connected to a specific arterial or venular tree. These cases can be identified with certainty as vessels, but their classification into arteries and veins is uncertain, even for an expert. Additionally, it is common to find crossings between arteries and veins in the retina. Thus, although for these positions either the artery or the vein is above the other, these  pixels can be regarded as simultaneously belonging to both types of vessels. 
This allows to account for continuous arterial and venular trees regardless of crossings below the other one.
The identification of these two special situations (uncertain vessels, and vessel crossings) is common in manually annotated retinal vessel classification datasets \cite{Hemelings:A/V:CMIG:2019,Hu:RITE:2013,Qureshi:ISCBMS:2013,Orlando:MICCAI:2018}. Figure~\ref{fig:example_img_dataset} shows an example of a ground truth image from the RITE dataset labelled this way, along with its decomposition into arteries, veins, crossings and uncertain vessels.


\subsection{Traditional approach}
\label{subsec:traditional_approach}

\begin{figure}
    \centering
    \includegraphics[width=0.84\textwidth]{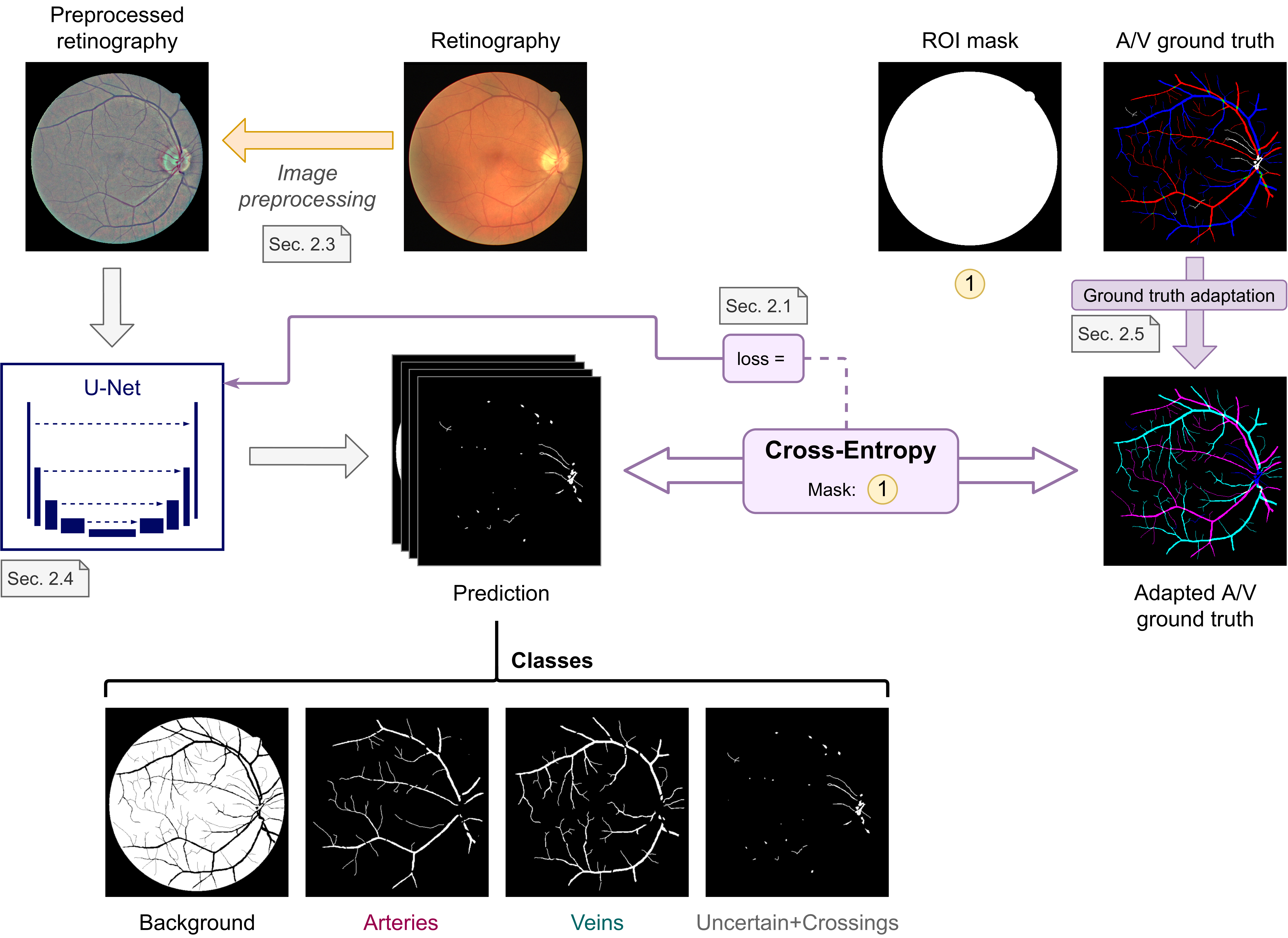}
    \caption{Traditional A/V segmentation and classification approach. The gray notes indicate the section of the article where the method is described.}
    \label{fig:traditional_approach}
\end{figure}

The traditional approach, depicted in Figure~\ref{fig:traditional_approach}, addresses the SSCAV as a single multi-class semantic segmentation task.
Thus, for each pixel, an FCNN predicts the likelihoods of $N$ mutually exclusive classes. 
To that end, the network output is composed of $N$ channels that are tied with each other using a softmax activation.
In this approach, Cross-Entropy (CE) is used as loss function, similarly to most state-of-the-art works in semantic segmentation. Formally, CE for joint vasculature segmentation and classification is defined as:

\begin{equation}
    \label{eq:CE}
    \mathcal{L}_{CE}\left(\textbf{f}(\textbf{r}),\textbf{s}\right) = -\sum_{\Omega}\sum_{c=1}^{N}\omega_c\cdot\textbf{s}_c \cdot \log\left(\textbf{f}(\textbf{r})_c\right) \ ,
\end{equation}

where $\textbf{f}(\textbf{r})_c$ denotes the network output for a given input retinography $\textbf{r}$ and class $c$, $\textbf{s}_c$ denotes its corresponding ground truth, $\omega_c$ denotes the weight of the class $c$, $N$ denotes the number of classes and $\Omega$ denotes a Region of Interest (ROI) mask, with the set of valid pixels in the image.

In this case, we consider $N=4$ classes, as depicted in Figure~\ref{fig:traditional_approach} (for this reason, we name this loss CE4). These classes correspond to ``background'', ``artery'', ``vein'', and ``uncertain or crossing''. Thus, as in \cite{Galdran:Uncertainty:ISBI:2019}, the fourth class includes both the vessels of uncertain class and the vessel crossings.

Following prior works \cite{Girard:Joint:AIM:2019,Hemelings:A/V:CMIG:2019,Galdran:Uncertainty:ISBI:2019}, we set the $\omega_c$ weights for the ``background'', ``artery'' and ``vein'' target classes to $1$, so that all pixels of these classes count equally for the loss. Conversely, the weights for the ``uncertain or crossing'' class are set to $0$.
Thus, neither the vessel crossings, which simultaneously belong to the artery and vein classes, nor the uncertain pixels, provide learning feedback in this approach. This is the most commonly used approach in previous works \cite{Xu:A/V:BOE:2018,Girard:Joint:AIM:2019,Hemelings:A/V:CMIG:2019,Ma:Multi-task:MICCAI:2019,Kang:A/V:CMPB:2020}. 

Using this classification approach, it is possible to obtain the vascular segmentation map by performing a pixel-wise addition of the predicted probability maps for arteries, veins and ``uncertain or crossings'', or using the inverse of the background class. However, it is not likely that the arterial and venular tree segmentation maps derived from the artery and vein classes are continuous, not even considering the ``uncertain or crossings'' class to complete them. The reason is that the network must decide to classify the vessel crossings as either of the above classes, in case of detecting them. Due to the fact that either the artery or the vein will be above the other, and the conditions for uncertain vessel classification are not commonly met at regular vessels, it is not probable that the ``uncertain or crossings'' class will be used. Thus the vascular tree for the vessel below will be discontinuous.

One possible solution to fix the vessel crossing discontinuities is to also use a weight $\omega_c=1$ for the ``uncertain or crossings'' class in Equation~\ref{eq:CE}, as in \cite{Galdran:Uncertainty:ISBI:2019}.  However, while this solution forces the network to mark the crossings as this specific class, there is no distinction among these regions and the uncertain vessels. Thus, an additional split into two classes, with a very low number of learning samples for each one, would be further necessary. In this work, we provide a more straightforward and balanced solution, as explained in the next section, that addresses this issue.


\subsection{Proposed multi-segmentation approach}
\label{subsec:MS_approach}

\begin{figure}
    \centering
    \includegraphics[width=\textwidth]{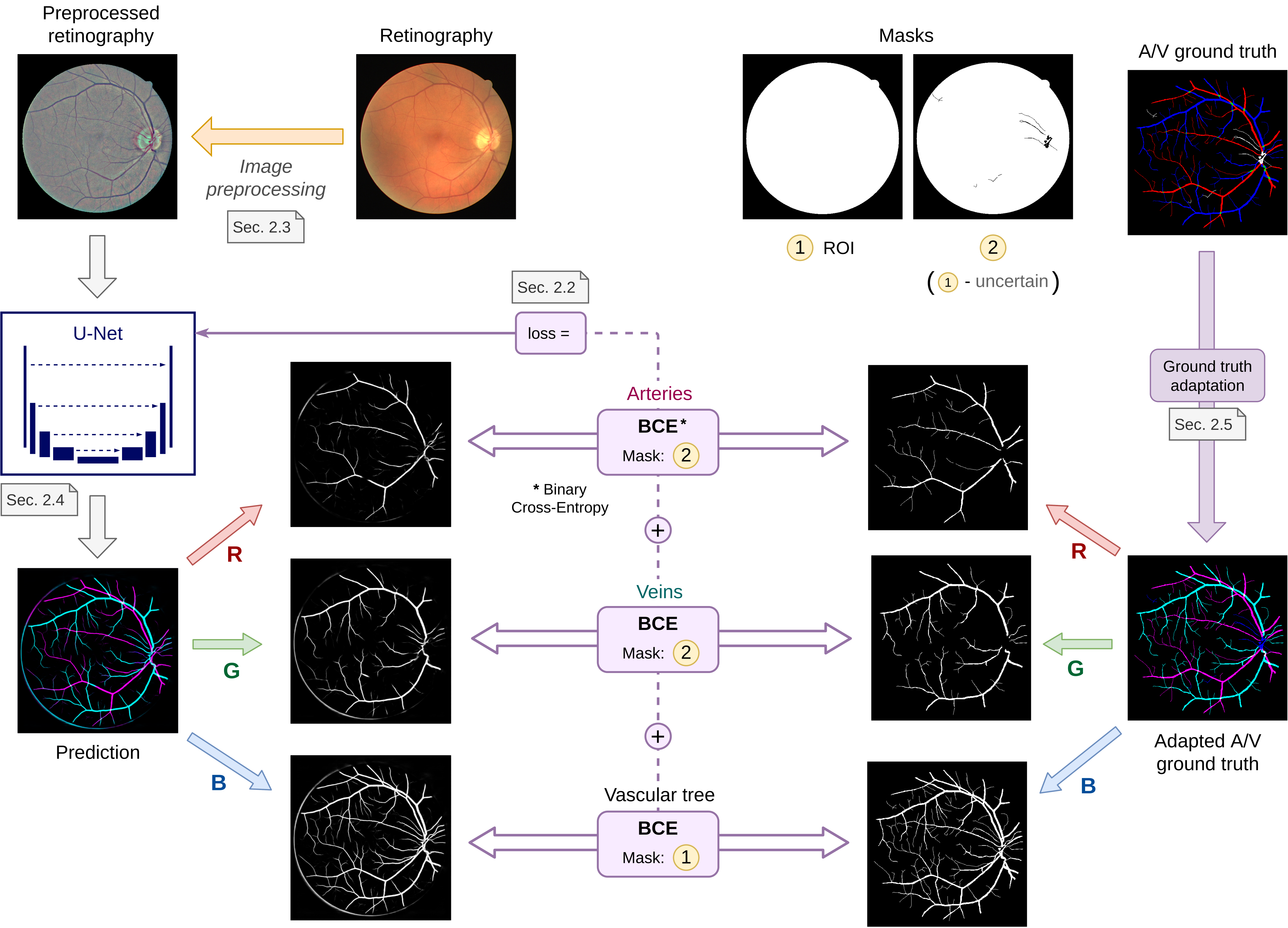}
    \caption{Proposed A/V segmentation and classification approach. In each case, the corresponding Binary Cross-Entropy (BCE) is calculated only for the pixels that match the region delimited by the mask. For vessels, the delimited region includes all the pixels of the ROI, while for arteries and veins, this region does not include ``uncertain'' vessels. The gray notes indicate the section of the article where the method is described.}
    \label{fig:ms_approach}
\end{figure}

Our proposed approach for SSCAV is depicted in Figure~\ref{fig:ms_approach}. 
To both segment and classify the retinal vessels, we decompose the SSCAV problem into three different tasks: arteries segmentation, veins segmentation and vessels segmentation. This multi-segmentation (MS) approach is in line with the dual nature of the retinal vascular tree, where arteries and veins individually spread throughout the retina and intersect frequently at different points. 
Moreover, following this approach, the three continuous vascular tree segmentation maps (arteries, veins and all the vessels) are also directly provided by the neural network.

In this case, the network output consists of 3 independent channels (see Figure~\ref{fig:ms_approach}). Each channel contains the predicted probability map for one of the structures of interest: arteries, veins and vessels. The total loss is computed as the sum of the individual losses for each structure, each being calculated as the Binary Cross-Entropy (BCE) between the manually annotated segmentation mask of the structure and the corresponding predicted probability map. Mathematically, the ``Binary Cross-Entropy by 3'' (BCE3) loss is defined as

\begin{equation}
\label{eq:BCE3}
    \mathcal{L}_{BCE3}\left(\textbf{f}(\textbf{r}),\textbf{s}\right) = -\sum_{c=1}^{3}\sum_{~\Omega_c}\left[\textbf{s}_c \cdot log\left(\textbf{f}(\textbf{r})_c\right)+\left(1-\textbf{s}_c\right) \cdot log\left(1-\textbf{f}(\textbf{r})_c\right)\right] \ ,
\end{equation}

where $\textbf{r}$ is the input retinography, $\textbf{f}(\textbf{r})_c$ and $\textbf{s}_c$ the network output and the ground truth for class $c$ and $\Omega_c$ is the set of all the pixels within a class-specific ROI mask. The class-specific ROI mask for the vessels class is the retinal ROI, and it is equivalent to the $\Omega$ used in Equation~\ref{eq:CE}. Differently, for the artery and vein classes, the ROI mask $\Omega_c$ is computed as the retinal ROI minus the vessel pixels of uncertain class, as labelled by the experts in the ground truth. Examples of these ROI masks are represented in Figure~\ref{fig:ms_approach}.

Notice that avoiding the uncertain pixels in the artery and vein segmentation losses (i.e. to mask them for the computation of the loss), and not in the vascular segmentation one, allows the system two things: first, to freely assign those pixels to artery or vein classes (as in the case of the traditional approach), and second, to simultaneously receive feedback for those pixels in the vascular segmentation subtask. Consequently, the network receives feedback for every pixel inside the ROI.

In addition to this feature, it is also relevant to note that, in this case, the ``uncertain'' class does not include the vessel crossings, unlike the ``uncertain or crossing'' class defined in the traditional approach.
Instead, in this approach, the vessel crossings are encoded as the superposition of the artery and vein segmentation maps, which should be both indicating positive class for these pixels. In order to do that, the ground truth $\textbf{s}_c$ for each class $c$ should be adapted to the output of the network and these restrictions. This can be achieved by considering the vessel crossing regions as positives in both artery and vein segmentation maps, and combining these two maps with the uncertain vessel regions to obtain the vascular tree segmentation map. An example of such an adapted ground truth is shown in Figure~\ref{fig:ms_approach}.
Additionally, this setting allows to detect vessel crossings through the intersection (product) of the predicted arteries and veins segmentation masks.


\subsection{Preprocessing}

Retinography datasets usually present a significant illumination and contrast variability among the images. This variability is due to different factors that affect the image acquisition. For example, distinct pupil sizes or exposure levels. In addition, retinographies also present illumination and contrast variability within the image due to the curvature of the retina or uneven lighting during the acquisition.

To correct this variability and therefore facilitate the training of the networks, most of the state-of-the-art-works implement an image preprocessing technique \cite{Xu:A/V:BOE:2018,Girard:Joint:AIM:2019,Hemelings:A/V:CMIG:2019,Galdran:Uncertainty:ISBI:2019,Ma:Multi-task:MICCAI:2019}. The ablation studies that are provided in these works clearly show the beneficial effect of the image preprocessing. However, the employed technique varies among the works, and there is not an accepted standard. In this context, and given the adequate results, we include the method proposed in the reference work of Girard et al. \cite{Girard:Joint:AIM:2019}. In particular, this technique performs a channel-wise global contrast enhancement and a local intensity normalization as follows:

\begin{equation}
    I^{C}_{\textnormal{norm}}=\sigma_0\frac{I^{C}-I^{C}_{l}}{\sigma_{I^{C}-I^{C}_{l}}} \ ,
\end{equation}

where $I^{C}_{\textnormal{norm}}$ is the normalized channel $C$, $I^{C}$ is the channel $C$ of the original retinography, $I^{C}_{l}$ is the low-pass filtered channel $C$ and $\sigma_{I^{C}-I^{C}_{l}}$ is the global standard deviation of the channel resulting from the subtraction of the low-pass filtered channel $I^{C}_{l}$ to the original channel $I^{C}$.

In our case, $\sigma_0 = 1$. Also, as low-pass filter, a gaussian filter with zero mean and standard deviation $\sigma = 10$ is used.

\begin{figure}
    \centering
    \begin{subfigure}{0.43\textwidth}
        \includegraphics[width=\linewidth]{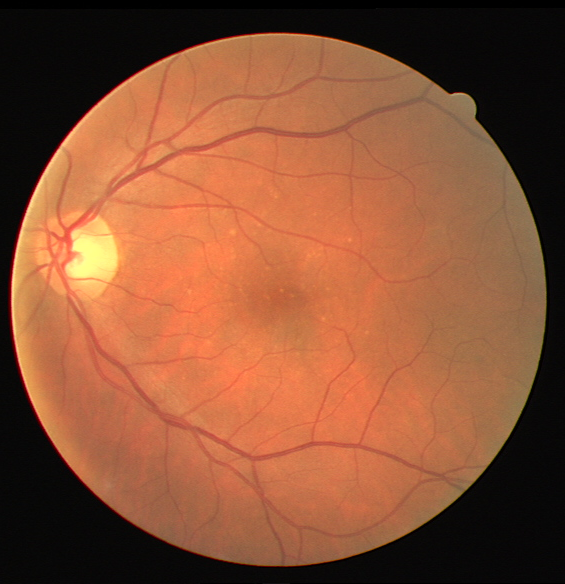}
        \caption{}
    \end{subfigure}
    \hspace{1cm}
    \begin{subfigure}{0.43\textwidth}
        \includegraphics[width=\linewidth]{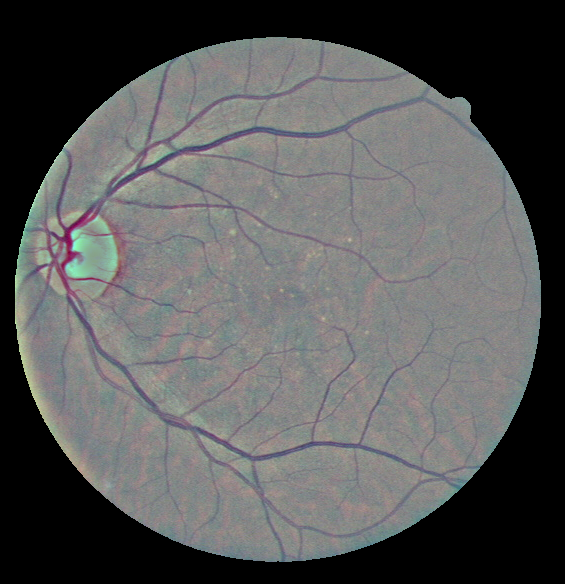}
        \caption{}
    \end{subfigure}\\~\\
    \begin{subfigure}{0.49\textwidth}
        \includegraphics[width=\linewidth]{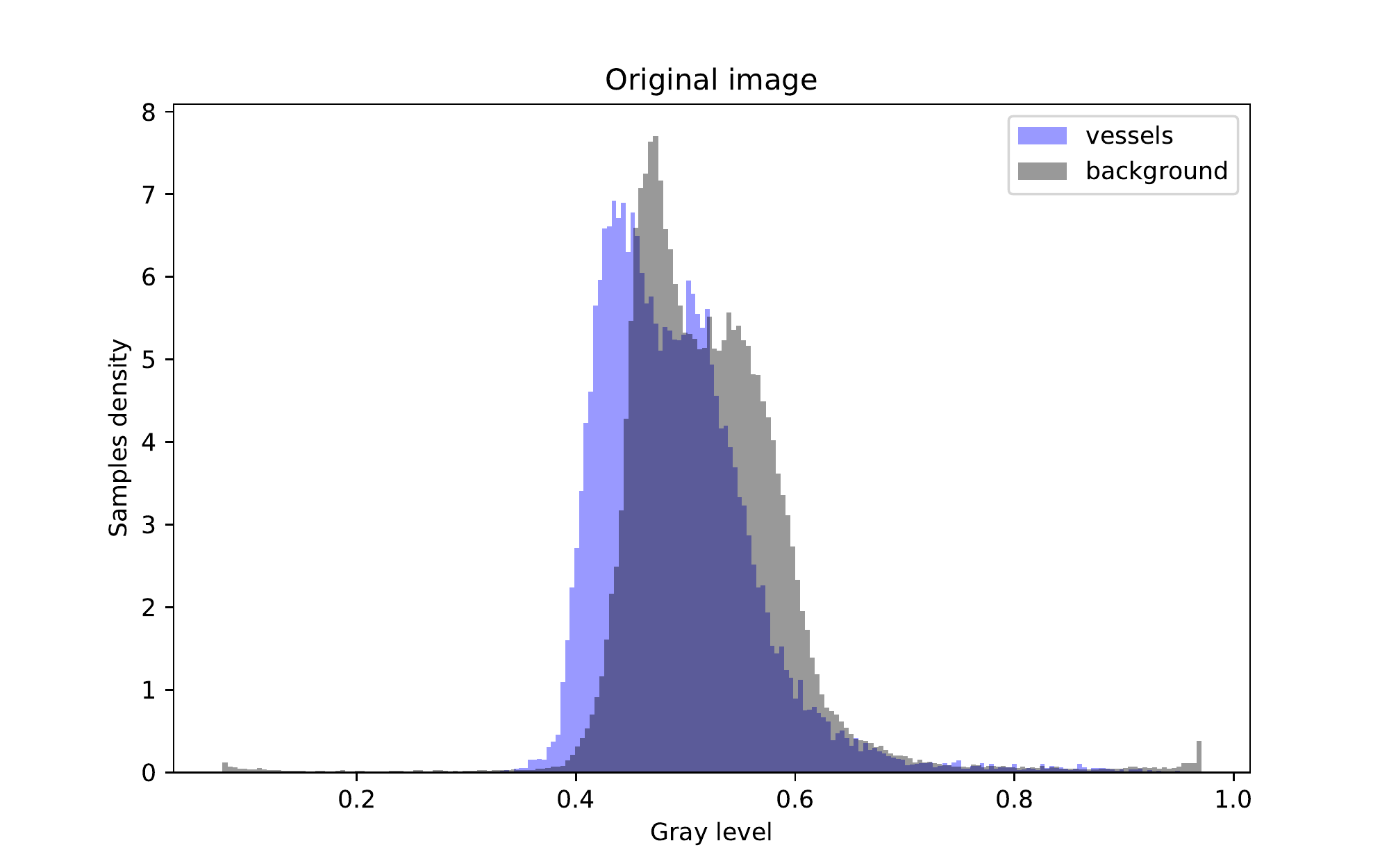}
        \caption{}
    \end{subfigure}
    \hfill
    \begin{subfigure}{0.49\textwidth}
        \includegraphics[width=\linewidth]{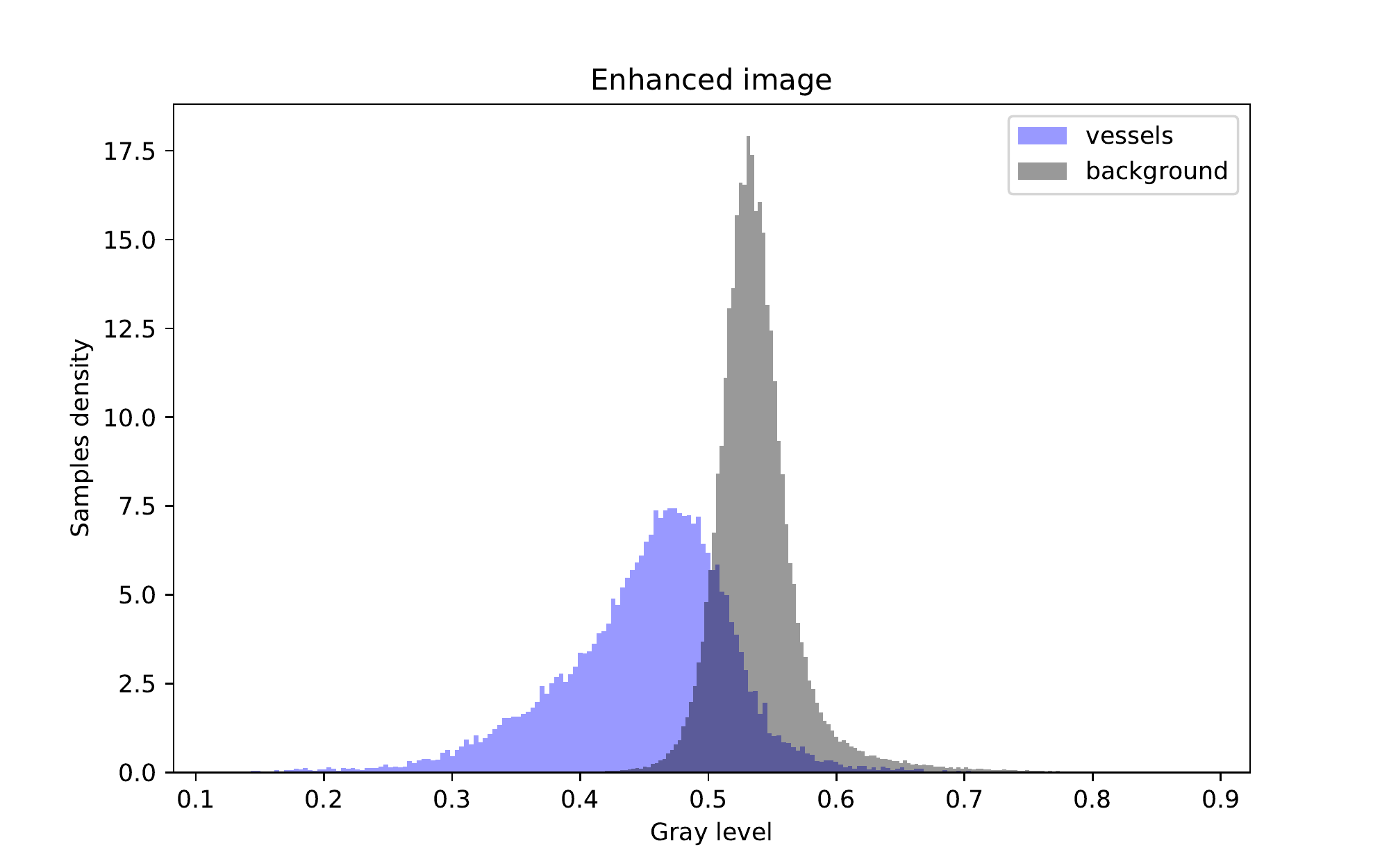}
        \caption{}
    \end{subfigure}
    \caption{Example of a RITE retinography before and after applying the preprocessing method, along with the gray level density histograms of vessel pixels and background pixels for each case. (a) Original retinography. (b) Preprocessed retinography. (c) Density histogram of the original retinography. (d) Density histogram of the preprocessed retinography.}
    \label{fig:preprocessing_example}
\end{figure}

In Figure~\ref{fig:preprocessing_example}, an example of a retinography before and after applying the preprocessing method is provided, along with the density histograms of vessel pixels and background pixels for each image. The histograms have been computed from the images converted to grayscale.
It can be seen at first glance that the enhanced image, i.e. the one that has been preprocessed, has a greater contrast between the vessels and the background and it presents a more uniform illumination. This enhancement can also be seen in the histograms of the images. 


\subsection{Network architecture}

For performing the SSCAV with both approaches, we use the fully convolutional neural network architecture \mbox{U-Net} \cite{Ronneberger:U-Net:MICCAI:2015}. This architecture was initially proposed for biomedical image segmentation, and its valuable results \cite{Zhu:NIPS:2017, Isola:CVPR:2017}, particularly in several tasks related to this field \cite{Falk:Cell:NM:2019,Fu:Optic:TMI:2018,Litjens:MIA:2017}, have made it to be considered a reference in computer vision. In fact, most of the state-of-the-art works in artery and vein classification are based on this architecture \cite{Galdran:Uncertainty:ISBI:2019,Girard:Joint:AIM:2019,Hemelings:A/V:CMIG:2019,Ma:Multi-task:MICCAI:2019,Xu:A/V:BOE:2018}. Furthermore, \mbox{U-Net} was successfully applied in our previous works for vasculature segmentation \cite{Morano:Segmentation:ECAI:2020, Hervella:ASOC:2020}, fovea localization, optic disc localization and segmentation \cite{Hervella:ASOC:2020}, and multimodal reconstruction \cite{Hervella:ESWA:2020}.

\begin{figure}
    \centering
    \includegraphics[width=0.99\textwidth]{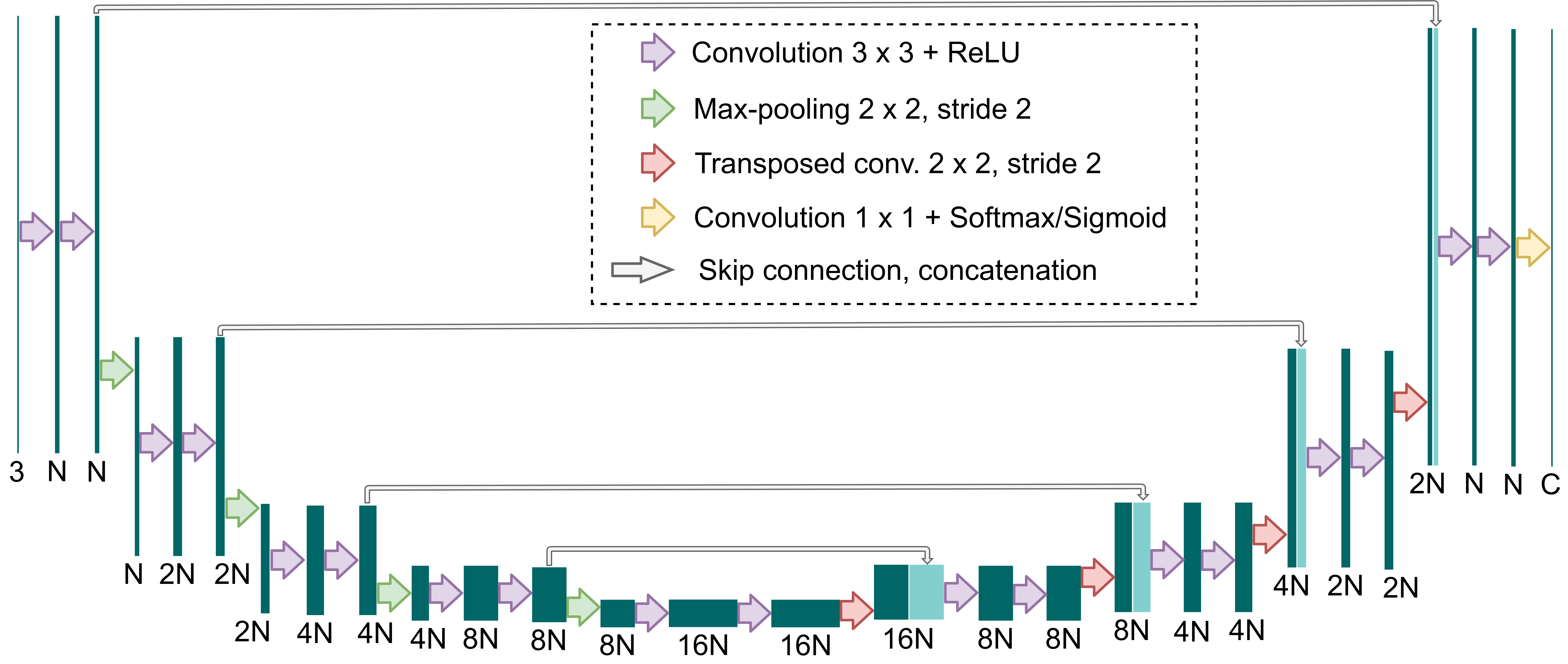}
    \caption{U-Net architecture. $N$ represents the number of base channels. In our case, $N=64$. $C$ stands for the number of channels of the output. In the MS approach, the sigmoid activation function is applied at the last convolutional layer of the network. In the traditional approach, this layer is followed by a softmax layer.}
    \label{fig:U-Net}
\end{figure}

In Figure~\ref{fig:U-Net}, a scheme of the \mbox{U-Net} architecture is shown. As can be seen in the figure, the network is composed of two main paths almost symmetrical: a contracting path (encoder), and an expansive path (decoder). This encoder-decoder shape is complemented with skip connections via concatenation between both paths.

The contracting path, or encoder, consists of 4 downsampling blocks. These blocks comprise two $3 \times 3$ convolutions, each followed by a ReLU, and a max-pooling of stride 2. Thus, after each downsampling block, the spatial resolution is halved. Moreover, the number of feature maps is duplicated.

The expansive path, or decoder, is composed of 4 upsampling blocks and, as said, it is almost symmetrical to the encoder. Each upsampling block consists of two $3 \times 3$ convolutions, each followed by a ReLU, and a $2 \times 2$ transposed convolution. These transposed convolutions produce the opposite effect to the max-pooling operations, duplicating the spatial resolution. Also, the number of feature maps is reduced by half. The result of each transposed convolution is then concatenated with the feature maps of the same spatial resolution from the contracting path through a skip connection. In this way, the features from the contracting path are transferred to the expanding path at different resolution levels, leading to more detailed results in the network output.

In the final part of the network, two $3 \times 3$ convolutions followed by a ReLU are applied to the result of the latter upsampling block, and then, a last convolution of $1 \times 1$ is used to reduce the number of feature maps as desired (e.g. to a certain number of classes). In the MS approach, a sigmoid activation function is applied at the last convolutional layer, while in the traditional approach, the last layer is followed by a softmax layer.


\subsection{Data}

For the experiments in this work, we employed the publicly available RITE dataset \cite{Hu:RITE:2013}, which is a reference standard for A/V classification. This dataset is based on the widely used Digital Retinal Images for Vessel Extraction (DRIVE) dataset \cite{Staal:DRIVE:2004}, which contains color retinography images and is likewise considered a reference standard for vascular segmentation. Both the DRIVE and RITE datasets are divided into the same training and test subsets, including 20 color fundus photographs each one. From these 40 images, 7 are from patients with diabetic retinopathy (mild early stage) and 33 are from healthy patients. All the images are $768 \times 584$ pixels with a circular Region Of Interest (ROI).

\begin{figure}
    \centering
    \begin{subfigure}{0.32\textwidth}
        \includegraphics[width=\linewidth]{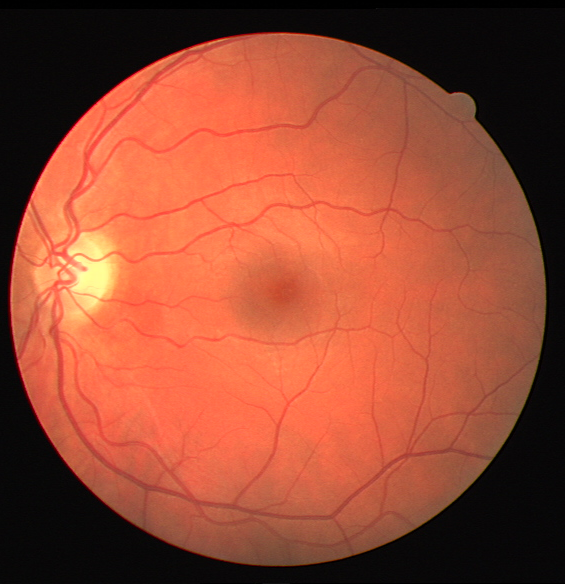}
        \caption{}
    \end{subfigure}
    \hspace*{\fill}
    \begin{subfigure}{0.32\textwidth}
        \includegraphics[width=\linewidth]{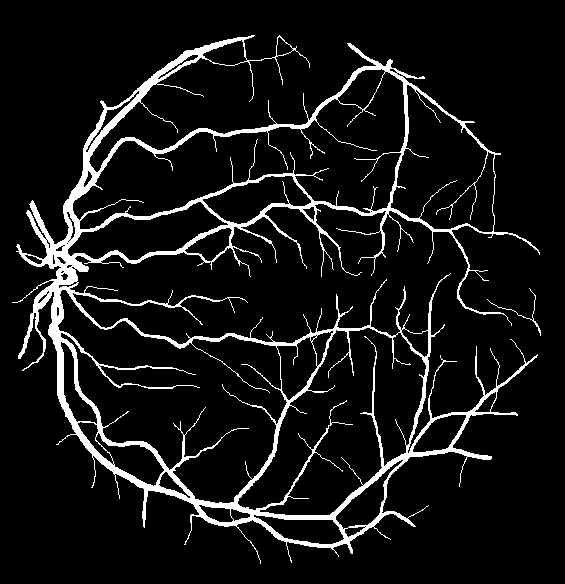}
        \caption{}
    \end{subfigure}
    \hspace*{\fill}
    \begin{subfigure}{0.32\textwidth}
        \includegraphics[width=\linewidth]{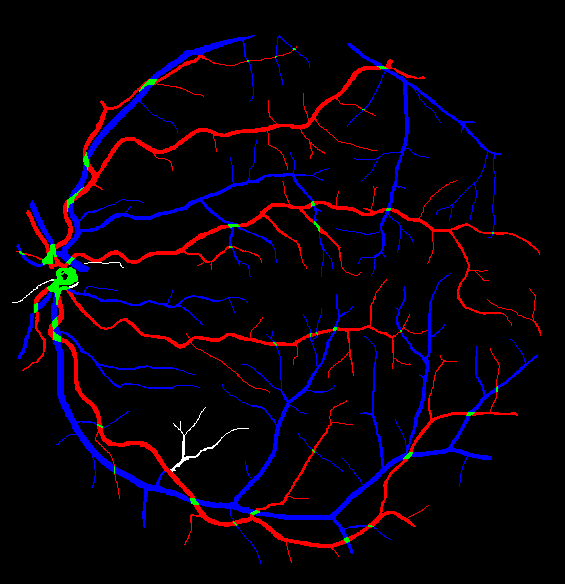}
        \caption{}
    \end{subfigure}
    \caption{Example of RITE retinography and its ground truths. (a) Retinography. (b) Vasculature segmentation ground truth. (c) A/V classification ground truth. In the A/V ground truth, arteries, veins, crossings and ``uncertain'' vessels are represented by the colors red, blue, green and white, respectively.}
    \label{fig:RITE_example}
\end{figure}

For all the images, the manual segmentation of the retinal vessels and their classification is available. The DRIVE dataset provides a gold standard manual vasculature segmentation map for each of the training and test images by an expert (first expert). Additionally, the test set images are annotated by another independent observer (second expert). 
The RITE dataset uses these manual annotations as reference to label the vessels. Specifically, the training set annotations are based on the first expert annotations from DRIVE, while the test set annotations are based on those of the second one. In the RITE annotations, all the vessels pixels were classified into one of four classes: \textit{artery}, \textit{vein}, \textit{crossing} or \textit{uncertain}. The \textit{crossing} class is used for pixels in the regions where arteries and veins overlap, while the \textit{uncertain} class is used for the pixels that the clinical experts have been able to identify as vessels but not to discriminate as arteries or veins. The distribution of data samples in these classes is shown in Table~\ref{tab:data_samples_distribution}.
\begin{table}
    \centering
    \caption{Distribution of data samples.}
    \label{tab:data_samples_distribution}
    
    \begin{tabular}{@{\extracolsep{4pt}}lrr}
    \toprule
        & Samples & \% \\
    \midrule
    Background & \numprint{3828898} & 87.52 \\
    \midrule
    Vessel & \numprint{546029} & 12.48 \\
    \tabitem Artery & \numprint{227210} & 5.19 \\
    \tabitem Vein & \numprint{278576} & 6.37 \\
    \tabitem Crossing & \numprint{14003} & 0.32 \\
    \tabitem Uncertain & \numprint{26240} & 0.60 \\
    \bottomrule
    \end{tabular}
\end{table}
Also, an example of a retinography and its corresponding vasculature segmentation and A/V classification ground truths is depicted in Figure~\ref{fig:RITE_example}.

For training the different networks, we use the 20 images of RITE-train with a random split of 15 and 5 images for training and validation, respectively. For testing, we use the entire RITE-test set.

\begin{figure}
    \centering
    \begin{subfigure}{0.32\textwidth}
        \includegraphics[width=\linewidth]{img/11_test_av.png}
        \caption{}
    \end{subfigure}
    \hspace*{\fill}
    \begin{subfigure}{0.32\textwidth}
        \includegraphics[width=\linewidth]{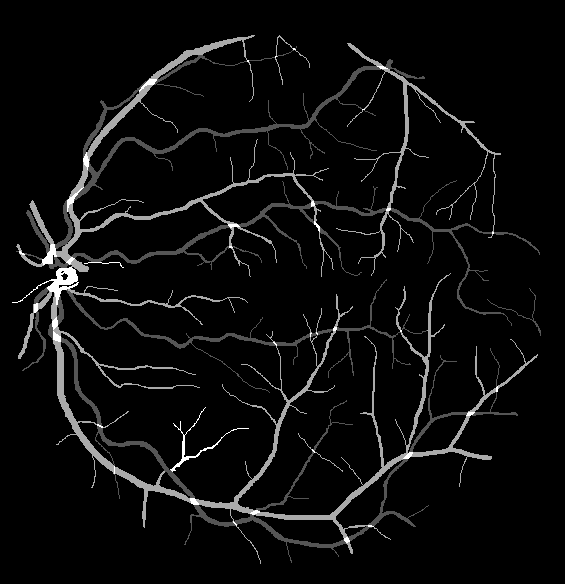}
        \caption{}
    \end{subfigure}
    \hspace*{\fill}
    \begin{subfigure}{0.32\textwidth}
        \includegraphics[width=\linewidth]{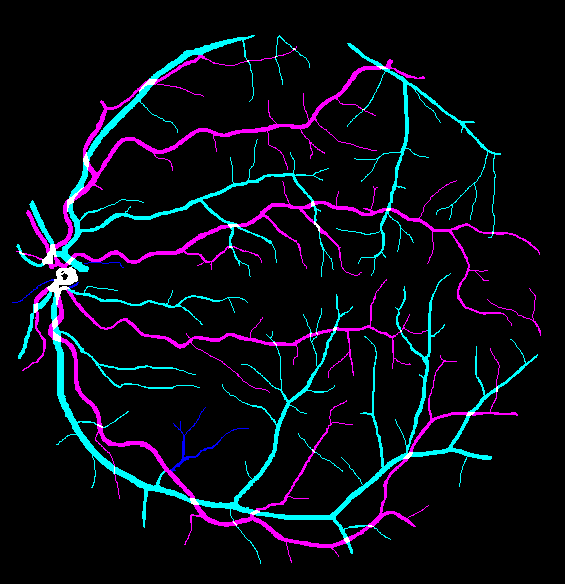}
        \caption{}
    \end{subfigure}
    \caption{Example of RITE retinography ground truth and its adaptations for the MS and traditional approaches. (a) Original RITE ground truth. (b) Adapted ground truth for the traditional approach. (c) Adapted ground truth for the MS approach.}
    \label{fig:RITE_example_adapted_GT}
\end{figure}

In order to train the networks following the traditional and MS approaches, the ground truth images from the RITE dataset are adapted, in each case, to the output of the network. In the traditional approach, each pixel of the ground truth is assigned to either background, artery, vein or ``uncertain or crossing'' class, which comprises both uncertain vessels and crossings. The resulting ground truth is a grayscale image in which each pixel has a numeric label of the class to which it belongs. On the other hand, in the MS approach, each pixel is assigned to any number of the following classes: artery, vein and vessel. In this case, crossings belong to the three classes, and uncertain vessels belong only to the ``vessel'' class. Background pixels are not assigned to any class. The resulting ground truth is an RGB image in which each channel contains a manual segmentation mask of one of the aforementioned structures. An example of a RITE ground truth adapted to both the traditional and the MS approaches can be found in Figure~\ref{fig:RITE_example_adapted_GT}.

Data processing for the evaluation of the different methods is described in detail in Section \ref{subsec:quantitative_evaluation}.


\subsection{Training details}

To train the different models, the Adam optimization algorithm is used~\cite{Kingma:Adam:2015}. The parameters of the algorithm were set as follows: initial learning rate $\alpha = 1 \times 10^{-4}$, decay rates $\beta_1 = 0.9$ and $\beta_2 = 0.999$. The training is performed at a constant learning rate, and early stopping is applied when the validation loss does not improve for 200 epochs. Both $\beta_1$ and $\beta_2$ values were taken from the original Adam paper \cite{Kingma:Adam:2015}, while the initial learning rate and the patience for early stopping were empirically established based on the learning curves for both training and validation data. To avoid an exhaustive exploration of the parameter space, we relied on prior evidence of the most suitable values that can be found in the literature \cite{Hemelings:A/V:CMIG:2019,Galdran:Uncertainty:ISBI:2019,Hervella:ASOC:2020,Hervella:ESWA:2020,Hervella:CMPB:2020}.

In all the cases, the parameters of the networks are initialized using the He et al.~\cite{He:Initialization:2015} initialization method with uniform distribution, as in \cite{Hervella:ASOC:2020,Hervella:CMPB:2020}.

As the training set contains only 15 images, we use online data augmentation to artificially increase the training samples. Thus, on each epoch, random transformations are applied to the input images.
These transformations consist of combinations of random color and intensity variations, slight affine transformations (rotation, scaling and shearing) along vertical and horizontal flipping. These transformations are the same as those used in \cite{Morano:Segmentation:ECAI:2020,Hervella:ESWA:2020}.

The proposed methodology is implemented in Python 3 using the open source framework PyTorch. Training and development was performed on a single NVIDIA Tesla P40 GPU with a memory size of 24 GB. The CPU was an Intel Xeon Gold 6146 CPU @ 3.20GHz, and the memory size, 16 GB. In such a system, the complete training of a U-Net network following the proposed methodology takes about 1.5 hours. In the test phase, it takes less than 0.01 seconds to segment one image using the GPU, and about 10 seconds using the CPU alone.


\subsection{Quantitative evaluation}
\label{subsec:quantitative_evaluation}

The quantitative evaluation of the presented approaches is performed by comparing the predicted vascular segmentation, as well as the identification of arteries and veins, against the manual ground truth annotations. As we are interested in both the segmentation and the classification, we divide the evaluation into three parts, focused on each one of these aspects and on the joint performance.
Also, only for the proposed approach, we include an extra evaluation focused on assessing the performance of the approach on the complex vessel crossings localization task.

The first part is focused on assessing the segmentation performance on the different structures of interest: arteries, veins and full vascular tree. For each structure, we compute the Receiver Operating Characteristic (ROC) and Precision-Recall curves, by considering each segmentation target as a binary classification problem against the background.
For the arteries and veins, we only take into account the pixels inside the ROI, excluding uncertain vessels and vessel crossings. The latter are excluded for performing a fair comparison with the traditional approach, which does not include vessel crossings in both arteries and veins probability maps. In each case, the positive class is the structure of interest: arteries, veins and vessels, and the negative class, everything else.

The ROC and PR curves are built applying a variable threshold to the predicted probability maps, so that it is possible to perform the evaluation without selecting an specific decision threshold. Also, to summarize these curves, we computed the Area Under the Curve (AUC) values in each case.

It is worth noting that we include the ROC and PR curves due to different reasons. The former is included due to its wide use in the literature as a default segmentation metric, while the latter is included due to its greater sensitiveness when the target classes are unbalanced \cite{Davis:ICML:2006}. In our case, for the segmentation of the different structures, the number of samples of the positive class, i.e. the vessels of each kind, is significantly lower than the number of samples from the negative class, which cover the whole retinal background and non-target structures. In this scenario, PR analysis is more convenient, as it presents a higher sensitiveness to changes in the number of false positives.

The second part of the evaluation is focused on assessing the performance of the approaches in artery/vein classification. This scenario represents the standard evaluation setting adopted by the state-of-the-art works for A/V classification. Thus, we follow the exact same evaluation procedure that is used in those works \cite{Girard:Joint:AIM:2019,Hemelings:A/V:CMIG:2019,Galdran:Uncertainty:ISBI:2019,Ma:Multi-task:MICCAI:2019,Kang:A/V:CMPB:2020}. Specifically, we build the confusion matrix for the artery and vein classes for the pixels that were annotated as ``vessel'' in the ground truth, excluding the uncertain vessels and the vessel crossings.
To that end, we consider the class with the highest probability at the network output.
We report Sensitivity, Specificity and Accuracy computed from this matrix, considering arteries as the positive class.
In this case, as with many prior works, we are reporting the results for the system at the operating point that results from the network training. 
Nevertheless, as this operating point highly depends on the training set and the used loss, the metrics derived from this confusion matrix are not suitable for comparing with systems at other operating points. 
Performing a threshold optimization to select other suitable operating point for this classification, as for example maximizing a balanced accuracy on the validation set \cite{Galdran:Uncertainty:ISBI:2019} also depends on the used validation set and metric. 
This does not improve the comparability as long as there is not a gold standard metric used as reference to evaluate the task.
Thus, in addition to reporting the performance of the selected system, we provide the ROC curves along with the AUC-ROC metric, considering a varying threshold for the artery-vein classification of the ground truth classified vessels.

Finally, in order to evaluate the joint artery-vein classification and the vascular segmentation, we also follow the common approach in some previous works. In this way, we provide the Sensitivity, Sensibility and Accuracy in the vessel-background classification using the trained system, along with the artery-vein classification performance metrics. However, this approach, although being adopted by the state-of-the-art works in A/V classification, is not providing a joint evaluation of artery-vein classification of the segmented vasculature. This evaluation includes all the pixels that were labelled as ``vessel'' in the ground truth, assuming that the predicted vessels masks are perfect. Nonetheless, this assumption is not true, as the models can only detect part of the total number of vessel pixels. In a real scenario, only these pixels would be classified into artery and vein classes, not all the pixels that actually belongs to vessels, as the standard evaluation assumes. Therefore, the standard evaluation does not truly reflects the overall performance of the models.

To overcome this issue, we include a different evaluation method, which constitutes the third part of the quantitative evaluation. This method is intended to determine, simultaneously, the performance of the approaches regarding the A/V classification and the vessels segmentation. For this end, we plot the vessel/background classification sensitivity against the A/V classification accuracy for multiple thresholds. In addition, to compute the latter for the different thresholds, only the detected vessel pixels labelled as such in the ground truth are taken into account (i.e. the intersection of detected and annotated vessels). The curve resulting from applying this evaluation clearly depicts the overall performance of the networks regarding the two involved tasks.

As commented above, we include an extra evaluation for the proposed approach. This evaluation is focused on assessing the performance of the approach in the vessels crossings localization task. For this purpose, it is necessary, first of all, to define how the output of the network and the ground truth are translated to a set of coordinates representing the vessel crossings. In this process, the first step is to obtain the ground truth and the predicted segmentation masks of the crossings. This is achieved by performing an element-wise product of the artery and vein channels, as the vessel crossings belong to both classes. Then, for the prediction, we use an intensity threshold followed by a small dilation to merge very close regions. Once the prediction is converted to binary and dilated, we perform a connected-component analysis (CCA) and compute the centroids of the detected regions for both the prediction and the ground truth. The coordinates of the detected centroids are the ones used in the evaluation. As gold standard, as it is done in \cite{Hervella:CMPB:2020}, we consider a predicted crossing a True Positive when it is within a certain distance $d$ of a crossing from the ground truth. Otherwise, it is considered a False Positive. Also, a crossing can be predicted only once. Thus, in each case, only the closest prediction within the distance $d$ is considered as a True Positive. The crossings that are not located within this distance of any ground truth crossing are considered False Negatives. This analysis is performed for a representative distance of 10 pixels ($d=10$). Finally, in order to build the PR curve, we use these measurements to compute the Precision and Recall metrics for a moving threshold over the entire test set.


\subsection{Experimental details}

To quantitatively evaluate the performance of the MS and the traditional approaches in the SSCAV task, we trained the same models using both approaches. Also, to evaluate the effect of the image preprocessing, we trained the networks using both the preprocessed images and the original ones. Thus, we compare a total of 4 alternatives. In addition, to take into account the stochasticity of the networks training, we performed 5 training repetitions with random initialization for each considered alternative.

Then, for the different alternatives, we performed the evaluation described in Section~\ref{subsec:quantitative_evaluation}. In each case, we built the mean ROC and PR curves (calculating its corresponding mean AUC values) and we computed all the selected metrics for the A/V classification (Sensitivity, Specificity and Accuracy) and its corresponding mean values. Furthermore, we performed the joint evaluation of the segmentation and the classification tasks. In each case, the standard deviation value is also included.

Ultimately, in order to determine if the results of the proposed approach are significantly better (statistically) than those of the traditional approach, we have performed a one-sided Student's $t$-test of the means of the AUC-ROC, AUC-PR and Accuracy metrics.


\section{Results and discussion}
\label{sec:results_discussion}

\begin{figure}
    \centering
    \includegraphics[width=0.9\textwidth]{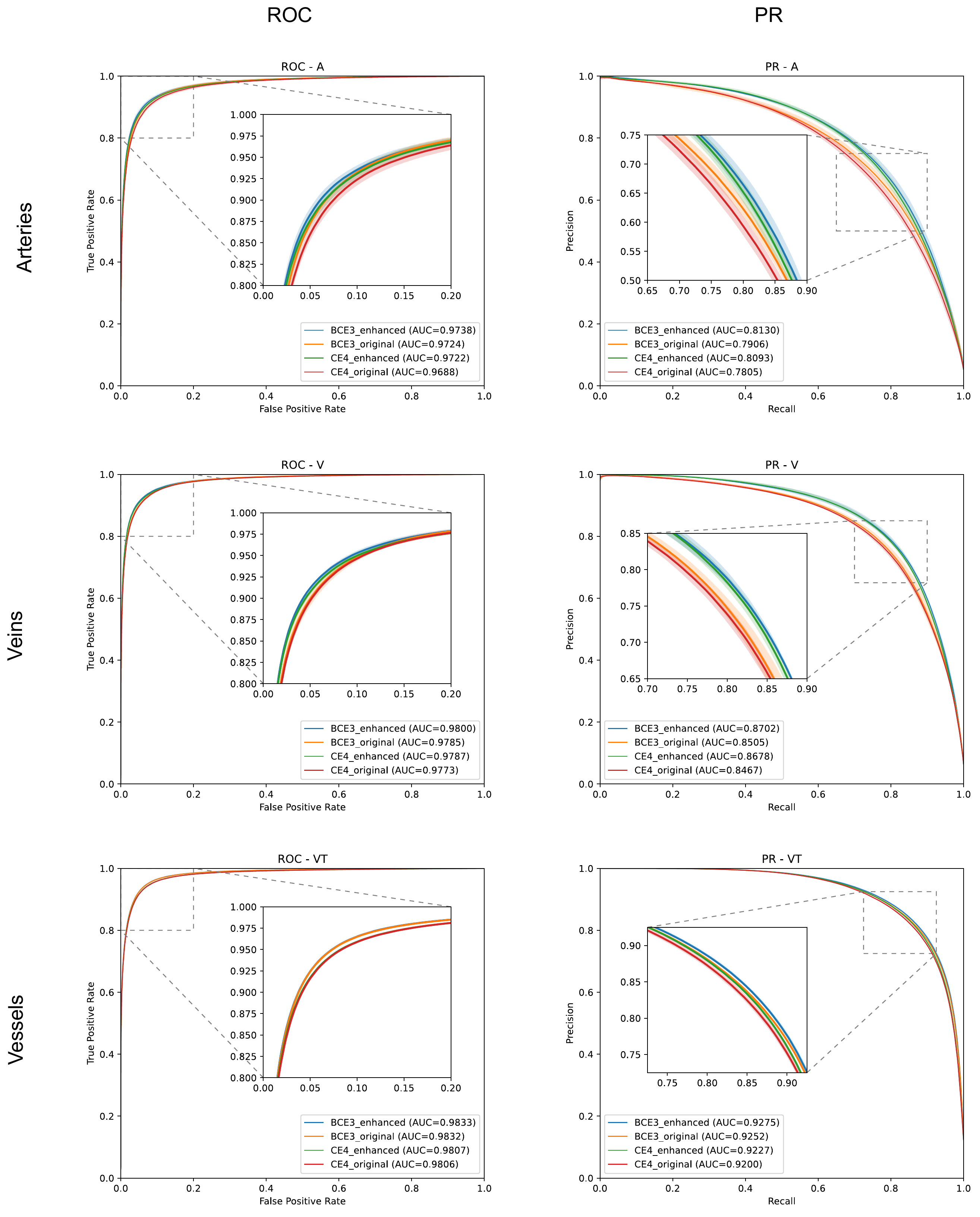}
    \caption{ROC and PR curves for the different structures in RITE-test for the networks trained using the MS and the traditional approaches and the enhanced and the original images.}
    \label{fig:segmentation_curves}
\end{figure}

\begin{figure}
    \centering
    \includegraphics[width=0.48\linewidth]{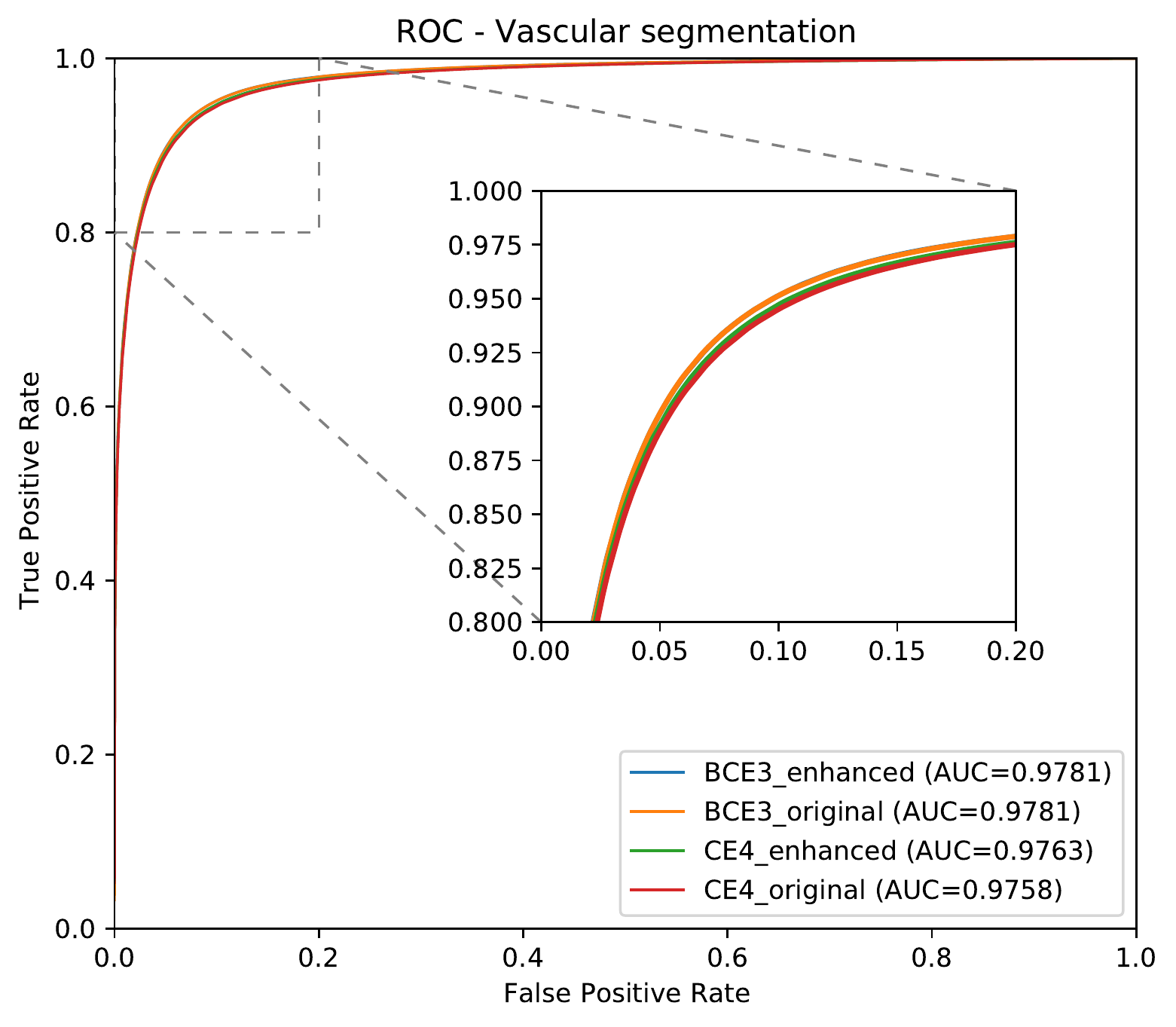}
    \hfill
    \includegraphics[width=0.48\linewidth]{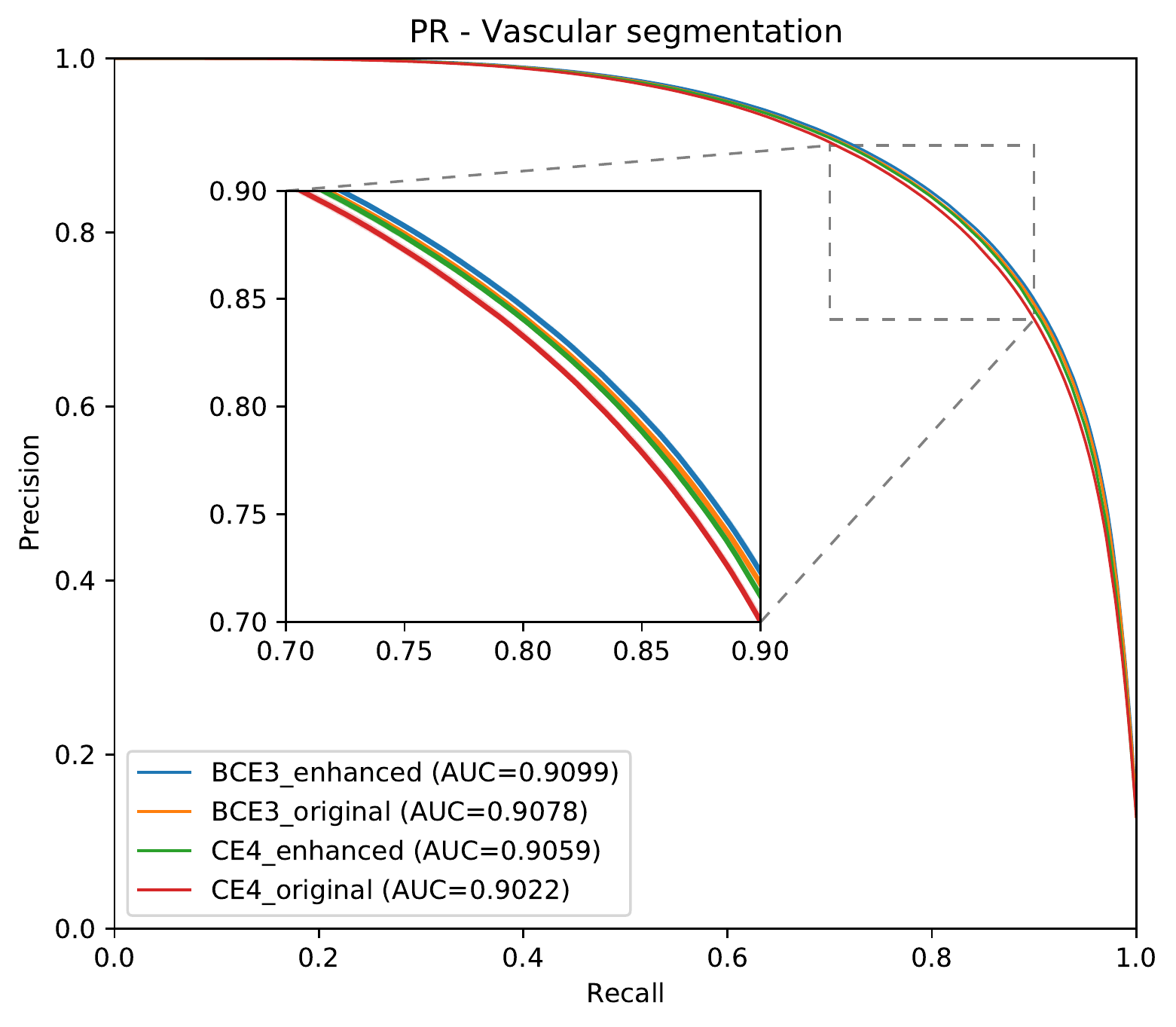}
    \caption{Vascular segmentation ROC and PR curves in DRIVE-test for the networks trained using the MS and the traditional approaches and the enhanced and the original images.}
    \label{fig:VO_curves_DRIVE}
\end{figure}

\begin{figure}
    \centering
    \includegraphics[width=0.48\linewidth]{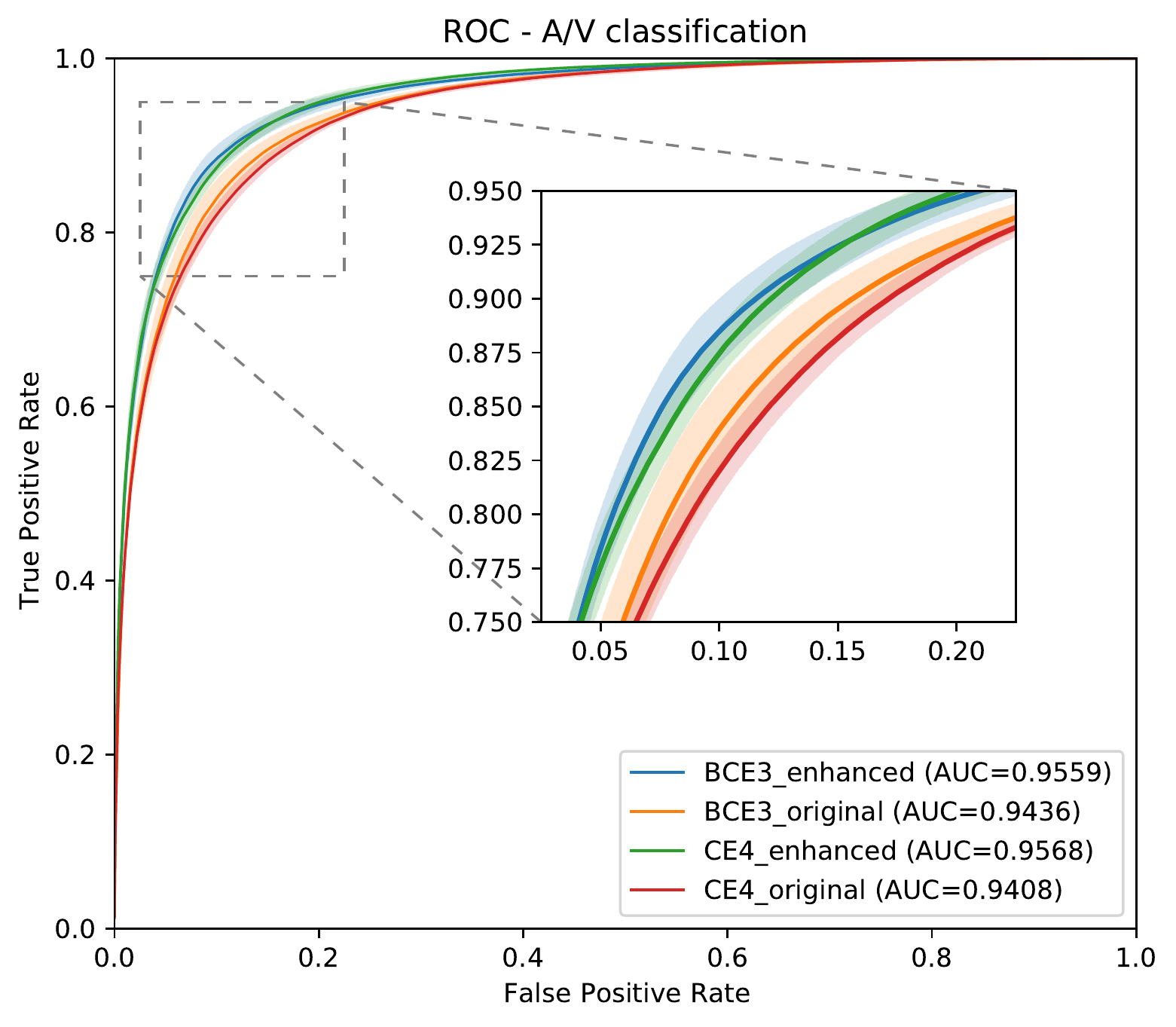}
    \hfill
    \includegraphics[width=0.48\linewidth]{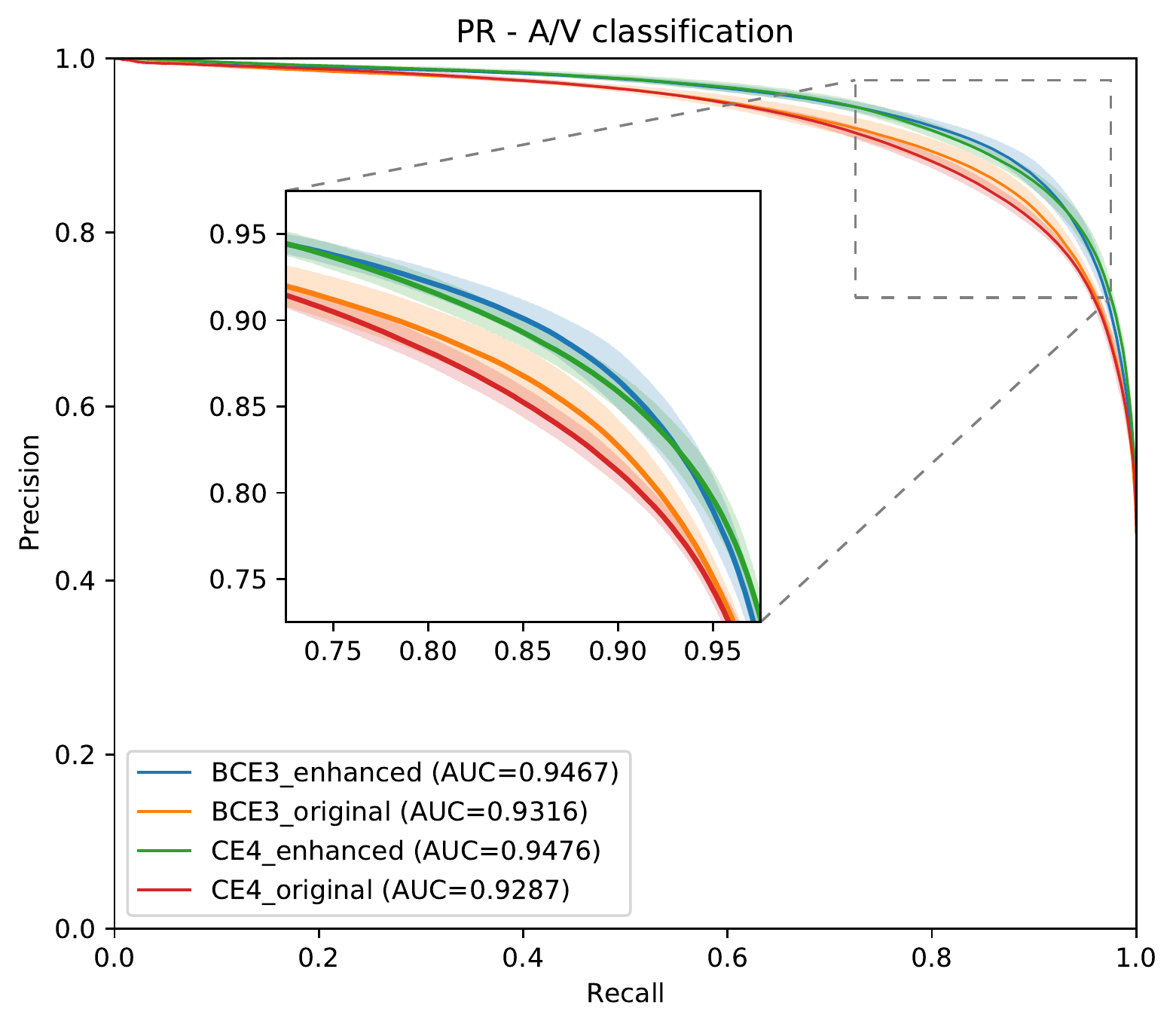}
    \caption{A/V classification ROC and PR curves in RITE-test for the networks trained using the MS and the traditional approaches and the enhanced and the original images.}
    \label{fig:curves_AVD}
\end{figure}

\begin{figure}
    \centering
    \includegraphics[width=0.6\textwidth]{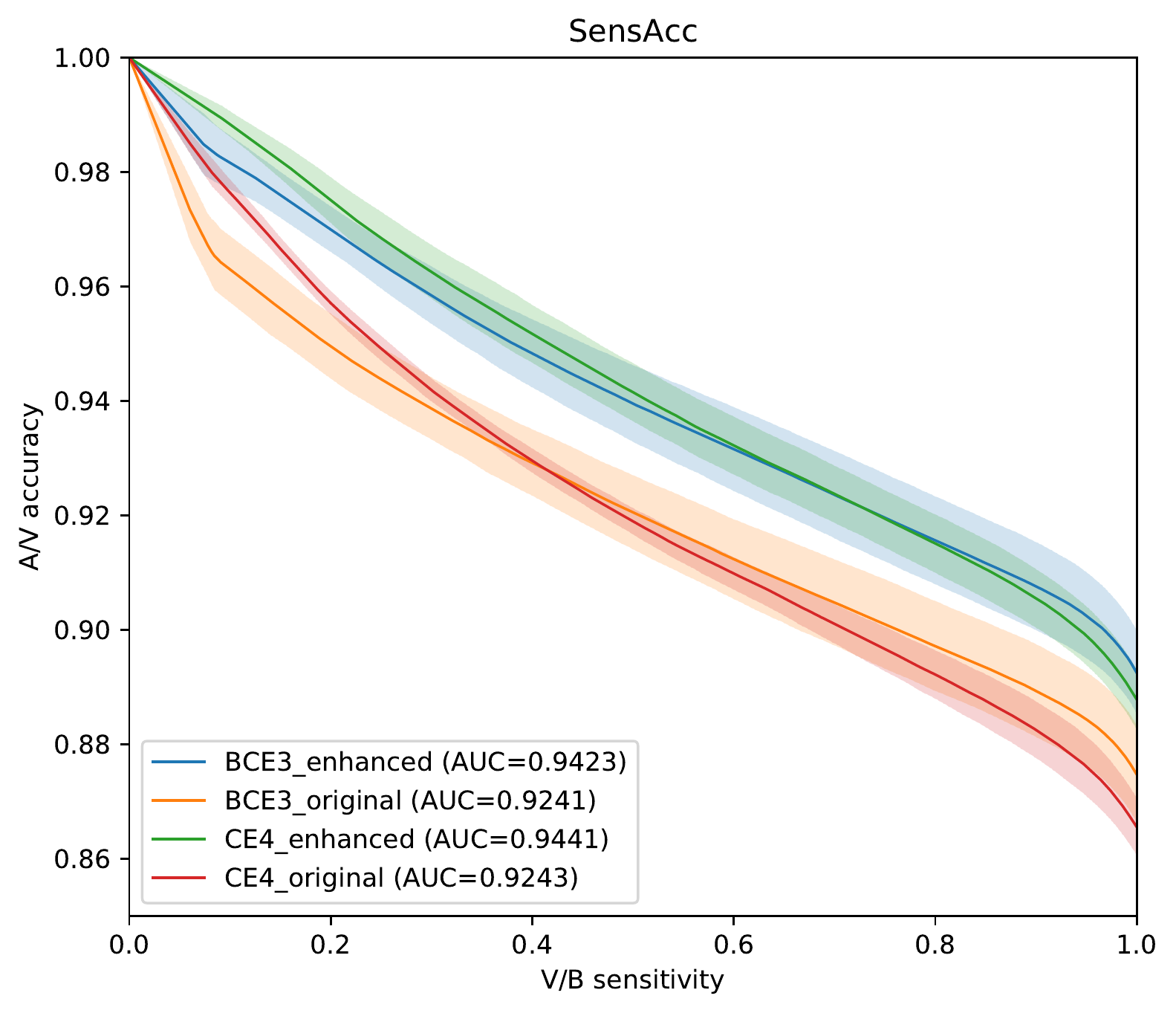}
    \caption{Vessels classification sensitivity against artery/vein classification accuracy in RITE-test for multiple thresholds and the different evaluated models.} 
    \label{fig:SensAcc_curves}
\end{figure}

\begin{figure}
    \centering
    \includegraphics[width=0.9\textwidth]{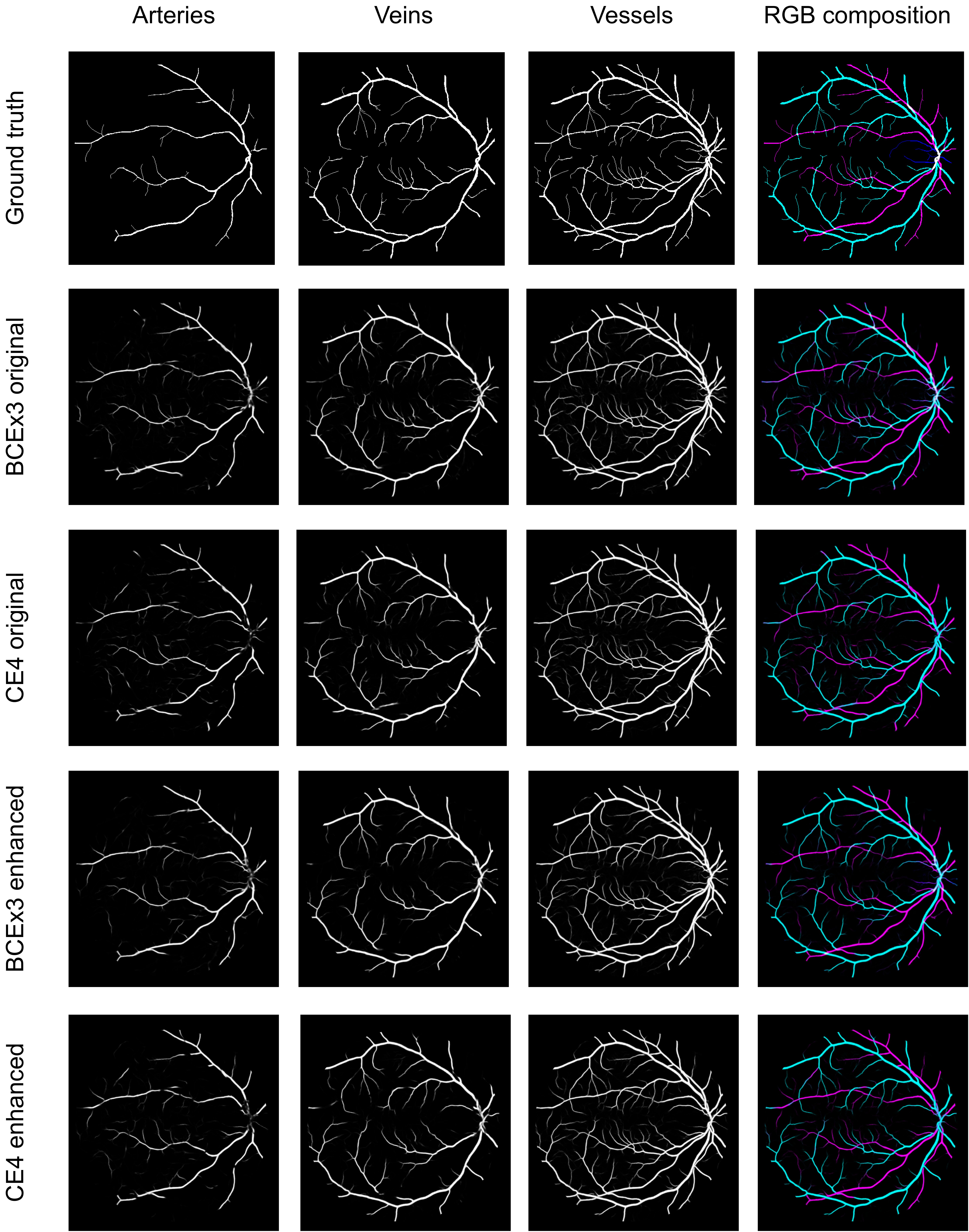}
    \caption{Examples of generated probability maps of the different classes and its RGB composition.}
    \label{fig:examples}
\end{figure}

In Figure~\ref{fig:segmentation_curves}, the mean ROC and PR curves in RITE-test for the networks that were trained using the BCE3 and CE4 losses with and without image preprocessing (``enhanced'' and ``original'', respectively) are depicted.
As commented in the previous sections, we include three ROC and PR curves, one for each structure of interest: arteries (A), veins (V), and Vascular Tree (VT)---that is, vessels---.
Figure~\ref{fig:VO_curves_DRIVE} depicts the ROC and PR curves for the different networks in the DRIVE-test set.
Additionally, all AUC values of the curves are summarized in Table~\ref{tab:segmentation_results}. 
\begin{table}
    \centering
    \caption{Segmentation results.}
    \label{tab:segmentation_results}
    \resizebox{\textwidth}{!}{%
    \begin{tabular}{@{\extracolsep{4pt}}lllllllll}
    \hline
    \multicolumn{1}{l}{Model} & \multicolumn{6}{l}{RITE-test} & \multicolumn{2}{l}{DRIVE-test} \\
    \cline{2-7}\cline{8-9}
     & \multicolumn{2}{l}{Arteries} & \multicolumn{2}{l}{Veins} & \multicolumn{2}{l}{Vessels} & \multicolumn{2}{l}{Vessels} \\
    \cline{2-3}\cline{4-5}\cline{6-7}\cline{8-9}
     & \multicolumn{1}{c}{AUC-ROC} & \multicolumn{1}{c}{AUC-PR} & \multicolumn{1}{c}{AUC-ROC} & \multicolumn{1}{c}{AUC-PR} & \multicolumn{1}{c}{AUC-ROC} & \multicolumn{1}{c}{AUC-PR} & \multicolumn{1}{c}{AUC-ROC} & \multicolumn{1}{c}{AUC-PR} \\
     & \multicolumn{1}{c}{(\%)} & \multicolumn{1}{c}{(\%)} & \multicolumn{1}{c}{(\%)} & \multicolumn{1}{c}{(\%)} & \multicolumn{1}{c}{(\%)} & \multicolumn{1}{c}{(\%)} & \multicolumn{1}{c}{(\%)} & \multicolumn{1}{c}{(\%)} \\ \hline

    \multicolumn{1}{l}{BCE3 original} & 97.24 $\pm$ 0.10 & 79.06 $\pm$ 0.77 & 97.85 $\pm$ 0.12 & 85.05 $\pm$ 0.44 & 98.32 $\pm$ 0.03 & 92.52 $\pm$ 0.08 & 97.81 $\pm$ 0.03 & 90.78 $\pm$ 0.09 \\ 

    \multicolumn{1}{l}{CE4 original} & 96.88 $\pm$ 0.32 & 78.05 $\pm$ 0.82 & 97.73 $\pm$ 0.12 & 84.67 $\pm$ 0.50 & 98.06 $\pm$ 0.06 & 92.00 $\pm$ 0.21 & 97.58 $\pm$ 0.04 & 90.22 $\pm$ 0.12 \\ 

    \multicolumn{1}{l}{BCE3 enhanced} & 97.38 $\pm$ 0.21 & 81.30 $\pm$ 0.73 & 98.00 $\pm$ 0.05 & 87.02 $\pm$ 0.38 & 98.33 $\pm$ 0.04 & 92.75 $\pm$ 0.12 & 97.81 $\pm$ 0.02 & 90.99 $\pm$ 0.07 \\ 

    \multicolumn{1}{l}{CE4 enhanced} & 97.22 $\pm$ 0.22 & 80.93 $\pm$ 0.89 & 97.87 $\pm$ 0.12 & 86.78 $\pm$ 0.54 & 98.07 $\pm$ 0.04 & 92.27 $\pm$ 0.08 & 97.63 $\pm$ 0.03 & 90.59 $\pm$ 0.07 \\ 
    
    \hline
    \end{tabular}%
    }
\end{table}
Complementing the segmentation results, Figure~\ref{fig:curves_AVD} depicts the ROC and PR curves of the different approaches for the A/V classification in the RITE-test set. These curves are built considering only the pixels labelled as ``vessel'' in the ground truth, being ``artery'' the positive class.
On the other hand, Table~\ref{tab:classification_results} reports the classification results in RITE-test for the different trained models.
\begin{table}
    \centering
    \caption{Classification results.}
    \label{tab:classification_results}
    \resizebox{\textwidth}{!}{%
    \begin{tabular}{@{\extracolsep{4pt}}lllllllll}
    \hline
    \multicolumn{1}{l}{Model} & \multicolumn{4}{l}{Artery/Vein} & \multicolumn{4}{l}{Vessel/Background} \\
    \cline{2-5}\cline{6-9}
    \multicolumn{1}{l}{} & Sens. (\%) & Spec. (\%) & Acc. (\%) & AUC-ROC (\%) & \multicolumn{1}{l}{Sens. (\%)} & \multicolumn{1}{l}{Spec. (\%)} & Acc. (\%) & AUC-ROC (\%) \\ \hline
    \multicolumn{1}{l}{BCE3 original} & 86.27 $\pm$ 1.53 & 88.64 $\pm$ 1.63 & 87.47 $\pm$ 0.85 &  94.36 $\pm$ 0.49& 80.26 $\pm$ 0.52 & 98.49 $\pm$ 0.07 & 96.23 $\pm$ 0.01 & 98.32 $\pm$ 0.03 \\
    \multicolumn{1}{l}{CE4 original} & 84.52 $\pm$ 1.53 & 88.46 $\pm$ 1.01 & 86.55 $\pm$ 0.50 & 94.08 $\pm$ 0.38& 77.52 $\pm$ 0.82 & 98.70 $\pm$ 0.12 & 96.08 $\pm$ 0.04 & 98.06 $\pm$ 0.06 \\
    \multicolumn{1}{l}{BCE3 enhanced} & 87.47 $\pm$ 2.09 & 90.89 $\pm$ 0.68 & 89.24 $\pm$ 0.73 & 95.59 $\pm$ 0.40 & 79.12 $\pm$ 1.22 & 98.65 $\pm$ 0.13 & 96.16 $\pm$ 0.05 & 98.33 $\pm$ 0.04 \\
    \multicolumn{1}{l}{CE4 enhanced} & 87.24 $\pm$ 1.24 & 90.26 $\pm$ 0.69 & 88.78 $\pm$ 0.53 & 95.68 $\pm$ 0.37 & 78.07 $\pm$ 1.64 & 98.67 $\pm$ 0.19 & 96.05 $\pm$ 0.05 & 98.07 $\pm$ 0.04 \\
    
    \hline

    \end{tabular}%
    }
\end{table}
Figure~\ref{fig:SensAcc_curves} depicts the vessels classification sensitivity against the artery/vein classification accuracy of the different models and multiple thresholds in RITE-test dataset.
\begin{figure}
    \centering
    \includegraphics[width=0.6\textwidth]{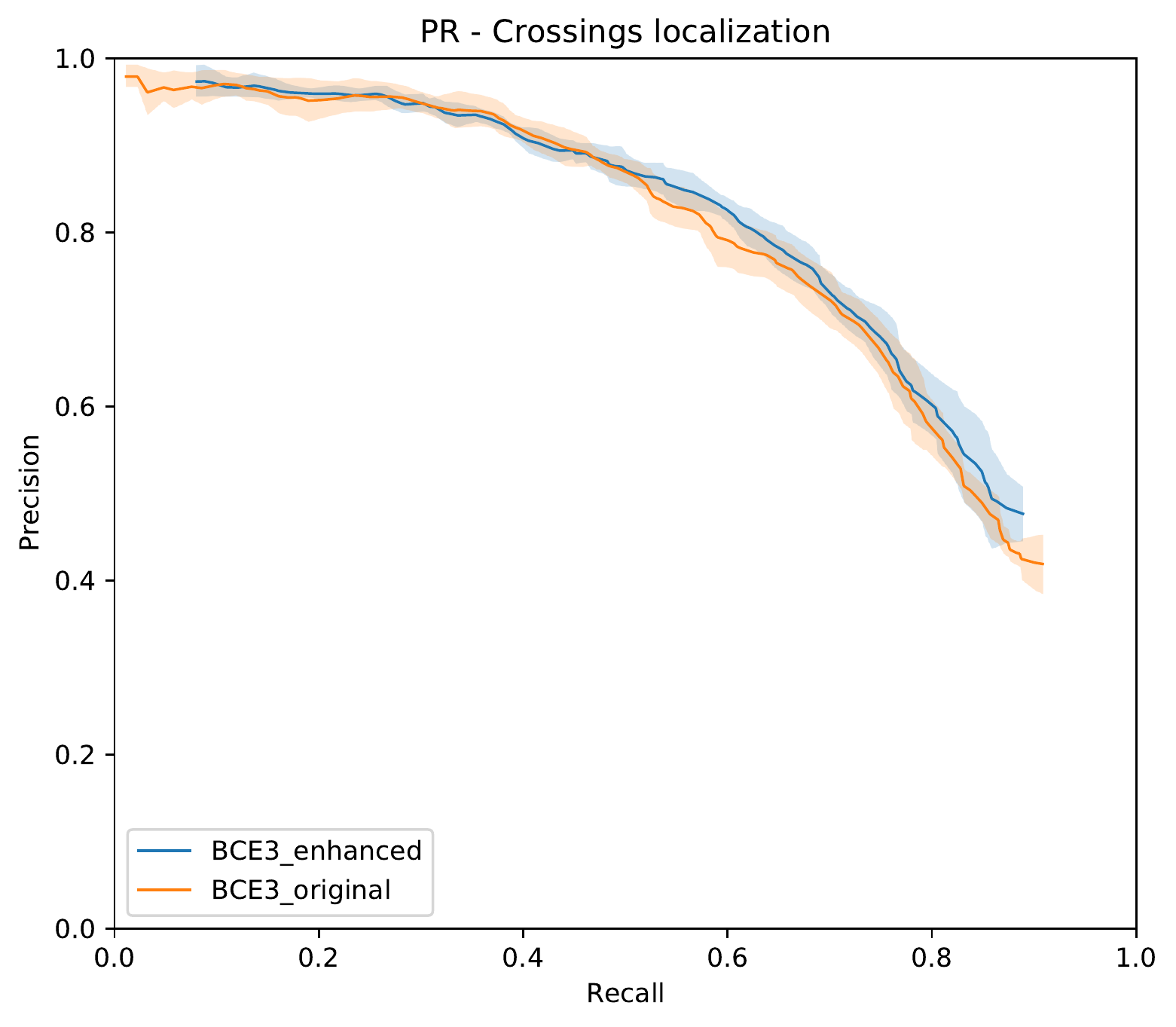}
    \caption{Vessels crossings localization PR curve in the RITE-test set for multiple thresholds and the different evaluated models trained using the MS approach.}
    \label{fig:PR_crossings_localization}
\end{figure}
Figure~\ref{fig:PR_crossings_localization} depicts the PR curve for vessel crossing localization in RITE-test set for the networks trained using the MS approach.
Lastly, Figure~\ref{fig:examples} shows representative examples of the predicted probability maps that were obtained by the trained models using both losses and the enhanced retinographies as input. In each case, the figure depicts the RGB composition of the predicted maps for a better clarity.

As it can be observed in the vasculature segmentation results of Figures~\ref{fig:segmentation_curves} and \ref{fig:VO_curves_DRIVE} as well as Table~\ref{tab:segmentation_results}, the best performance is achieved by the proposed MS approach using the BCE3 loss, either using the enhanced or the original retinography as input, and regardless of the target class. 
For each structure of interest, i.e. arteries, veins and vessels, both the AUC-ROC and AUC-PR values are higher with the MS approach than with the traditional CE4 one. Furthermore, AUC-PR and AUC-ROC for vessels are significantly higher for the MS approach than for CE4 ($p<0.001$), as well as other statistics like AUC-ROC for arteries ($p<0.05$) and AUC-ROC por veins ($p<0.05$).

\begin{figure}
    \centering
    \includegraphics[width=\textwidth]{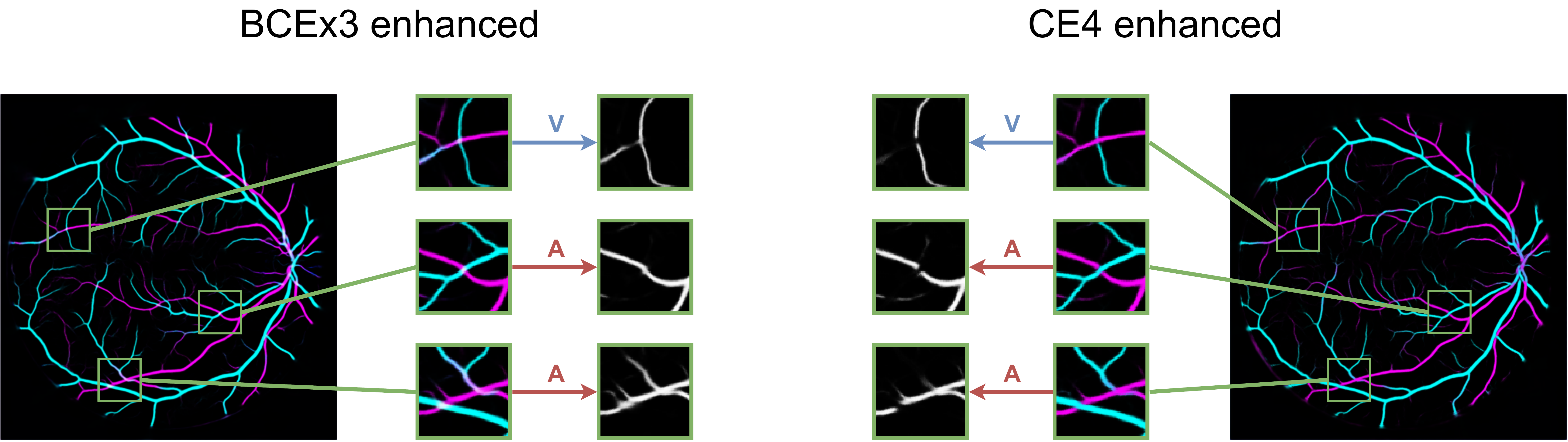}
    \caption{Examples of arteries, veins and vessels probability maps (RGB compositions) generated by the models trained using the MS and the traditional approaches with preprocessed images.}
    \label{fig:examples_zoom}
\end{figure}

In addition to the higher segmentation performance using BCE3, crossings are handled in a more intuitive way. In this case, crossing pixels are simply assigned to both artery and vein classes at a time and the network is able to detect them (see PR curve depicted in Figure~\ref{fig:PR_crossings_localization}) while allowing to achieve a continuous segmentation of both the arterial and venular trees. For CE4, differently, the crossings are mostly treated as a separate class (along with the uncertain vessels), either to detect them \cite{Galdran:Uncertainty:ISBI:2019}, or to let the network detect the artery or the vein above the other~\cite{Xu:A/V:BOE:2018,Girard:Joint:AIM:2019,Ma:Multi-task:MICCAI:2019,Kang:A/V:CMPB:2020}. There is also some approach in which the crossings belong to the same class as that of the upper vessel \cite{Hemelings:A/V:CMIG:2019}. Nevertheless, whichever of these alternatives gives raise to incomplete segmentation maps for both arteries and veins. Multiple examples of this effect can be found in Figure~\ref{fig:examples_zoom}. Taking all this into account, it can be stated that the MS strategy favors a better segmentation of the different structures, and handles the different cases in a much simpler way.

Regarding the artery/vein discrimination, although the MS alternatives achieve highly positive results (see Table~\ref{tab:classification_results}), it cannot be affirmed that they perform better than the traditional alternatives. Considering the variability, the A/V classification results of both approaches are not significantly different.
However, for the vessel/background discrimination, the situation is different. In this case, consistently with the higher AUC values for vessels segmentation, the MS approach achieves significantly better results ($p<0.01$) for both Accuracy and AUC-ROC. As can be seen in Table~\ref{tab:classification_results}, when training with the original images, the sensitivity values of the MS approach are, on average, more than 2.5\% higher than those of the traditional approach. Similarly, when training with the enhanced images, the improvement is about 1\%. We focus on sensitivity, at a similar level of specificity, due to the positive class that is ``vessel'', and the sensitivity measures the proportion of positives that were correctly identified. Specificity and accuracy, although they are also relevant, are much less sensitive in this scenario, since they take into account the background pixels, much more numerous and easier to classify.

The results depicted in Figure~\ref{fig:SensAcc_curves} are also in this line. Both methods, the traditional CE4 loss and the proposed MS with BCE3 loss, perform similarly when evaluated in simultaneous vascular segmentation and artery/vein classification. It is remarkable, however, that for a high vascular segmentation sensitivity the proposed approach achieves a slightly higher artery/vein classification accuracy.

In light of all these results together, it can be stated that the MS approach represents the more convenient alternative, since it is able to detect more vessels with similar A/V classification accuracy.

Another relevant point that can be observed in the different figures is the beneficial effect of the preprocessing for the A/V classification. As can be seen in Table~\ref{tab:classification_results} and Figure~\ref{fig:SensAcc_curves}, the performance of the networks is, in general, better when they are trained using the enhanced images than when they are trained using the original ones. In fact, this improvement can be clearly appreciated in Figure~\ref{fig:examples}, which depicts the probability maps that were predicted by the networks for a RITE-test retinography. When comparing the predictions of the ``enhanced'' and the ``original'' alternatives, it can be observed that some manifest classification errors disappear, an thus the arteries and veins continuity is slightly better preserved. Regarding the vessels segmentation, in light of the results, the preprocessing does not lead to such a significant improvement in the networks performance.


\subsection{Comparison with the state of the art}

Table~\ref{tab:classification_SOTA} shows the comparison with the state-of-the-art works for the joint vessel segmentation and the A/V classification tasks.
\begin{table}
    \centering
    \caption{Comparison with the state of the art for the joint vessels segmentation and Artery/Vein classification task in the RITE dataset.}
    \label{tab:classification_SOTA}
    \resizebox{\textwidth}{!}{%
    \begin{tabular}{@{\extracolsep{4pt}}lllllllll}
    \hline
    \multicolumn{1}{l}{Method} & Year & \multicolumn{3}{l}{Artery/Vein classification} & \multicolumn{4}{l}{Vessels segmentation} \\
    \cline{3-5} \cline{6-9}
    &  & Sens. (\%) & Spec. (\%) & Acc. (\%) & Sens. (\%) & Spec. (\%) & Acc. (\%) & AUC-ROC (\%) \\ \hline
    
    \multicolumn{1}{l}{Girard et al. \cite{Girard:Joint:AIM:2019}} & 2019 & 86.3 & 86.6 & 86.5 & 78.4 & 98.1 & 95.7 & 97.2 \\

    \multicolumn{1}{l}{Galdran et al. \cite{Galdran:Uncertainty:ISBI:2019}} & 2019 & 89 & 90 & 89 & \textbf{94} & 93 & 93 & 95 \\

    \multicolumn{1}{l}{Kang et al. \cite{Kang:A/V:CMPB:2020}} & 2020 & 88.63 & 92.72 & 90.81 & - & - & - & - \\
    
    \multicolumn{1}{l}{Ma et al. \cite{Ma:Multi-task:MICCAI:2019}} & 2019 & \textbf{92.2} & \textbf{93.0} & 92.6 & 76.16 & 98.11 & 95.70 & 98.10 \\

    \multicolumn{1}{l}{Hemelings et al. \cite{Hemelings:A/V:CMIG:2019}} & 2019 & - & - & \textbf{94.87} & - & - & - & - \\

    \multicolumn{1}{l}{Proposed} & 2020 & 87.47 $\pm$ 2.09 & 90.89 $\pm$ 0.68 & 89.24 $\pm$ 0.73 & 79.12 $\pm$ 1.22 & \textbf{98.65 $\pm$ 0.13} & \textbf{96.16 $\pm$ 0.05} & \textbf{98.33 $\pm$ 0.04} \\ \hline
    
    \end{tabular}%
    }
\end{table}
In this comparison, we include the results that were reported by relevant methods that have been evaluated on the publicly available RITE-test set.
However, some of the works do not evaluate the vessel segmentation performance of their methods \cite{Hemelings:A/V:CMIG:2019,Kang:A/V:CMPB:2020}, although the proposed methods aim at performing the vascular segmentation. 
As previously stated, the evaluation of the artery/vein classification task in the state-of-the-art works, and herein adopted for comparison, considers the ground truth vessels. 
However, this evaluation does not take into account if the networks are not able to detect all the vessels, which is relevant as in a real case scenario only the detected vessel pixels would be classified into artery and vein. In this way, if a model presents a high performance on the A/V classification, but is not able to detect most of the vessels, its overall performance is low. Consequently, an appropriate evaluation of the SSCAV should include not only an assessment of how well the model classifies the vessels, but also an assessment of how well the model detects those vessels.

\begin{figure}
    \centering
    \begin{subfigure}{0.48\textwidth}
        \includegraphics[width=\linewidth]{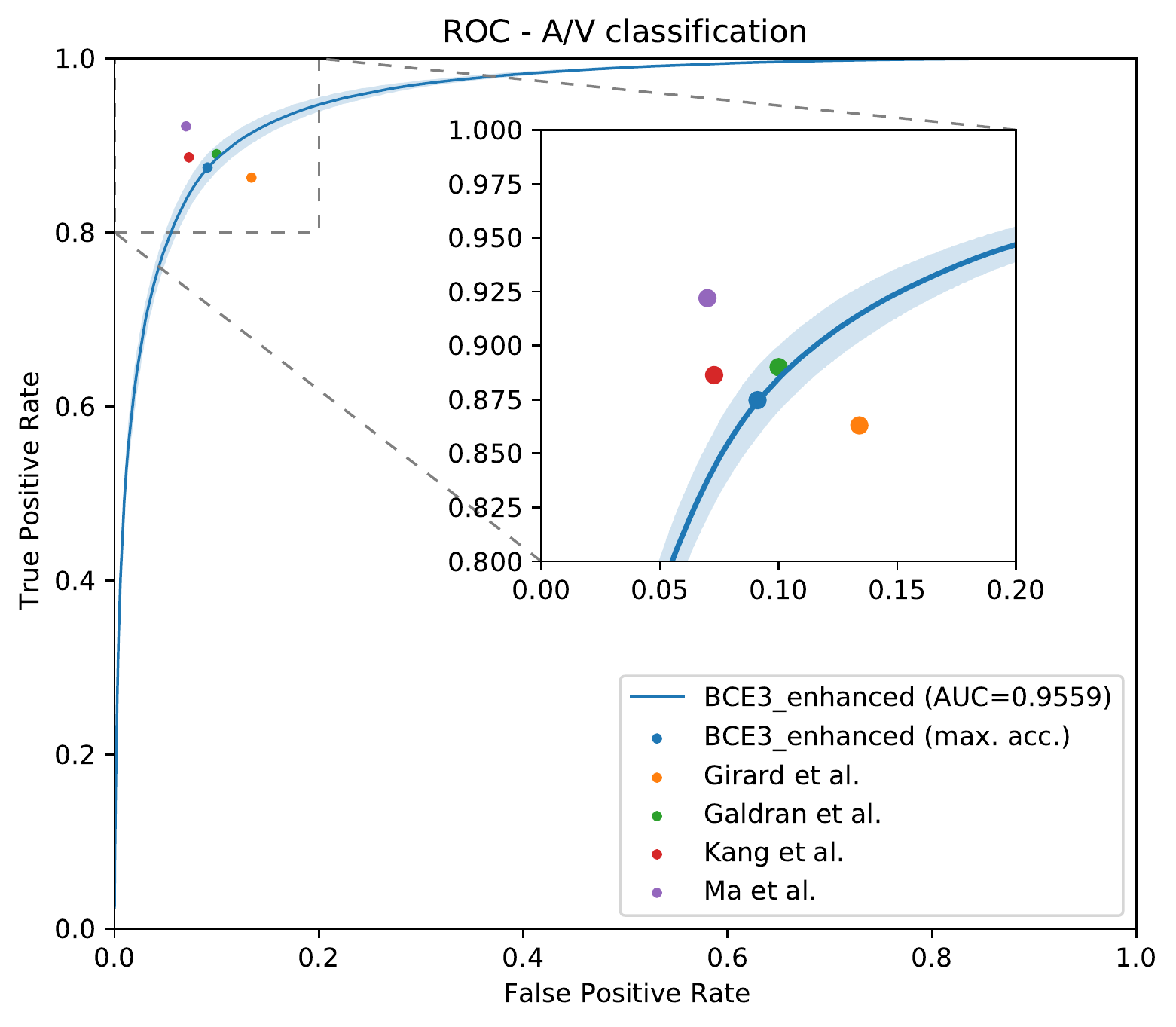}
        \caption{}
    \end{subfigure}
    \hfill
    \begin{subfigure}{0.48\textwidth}
        \includegraphics[width=\linewidth]{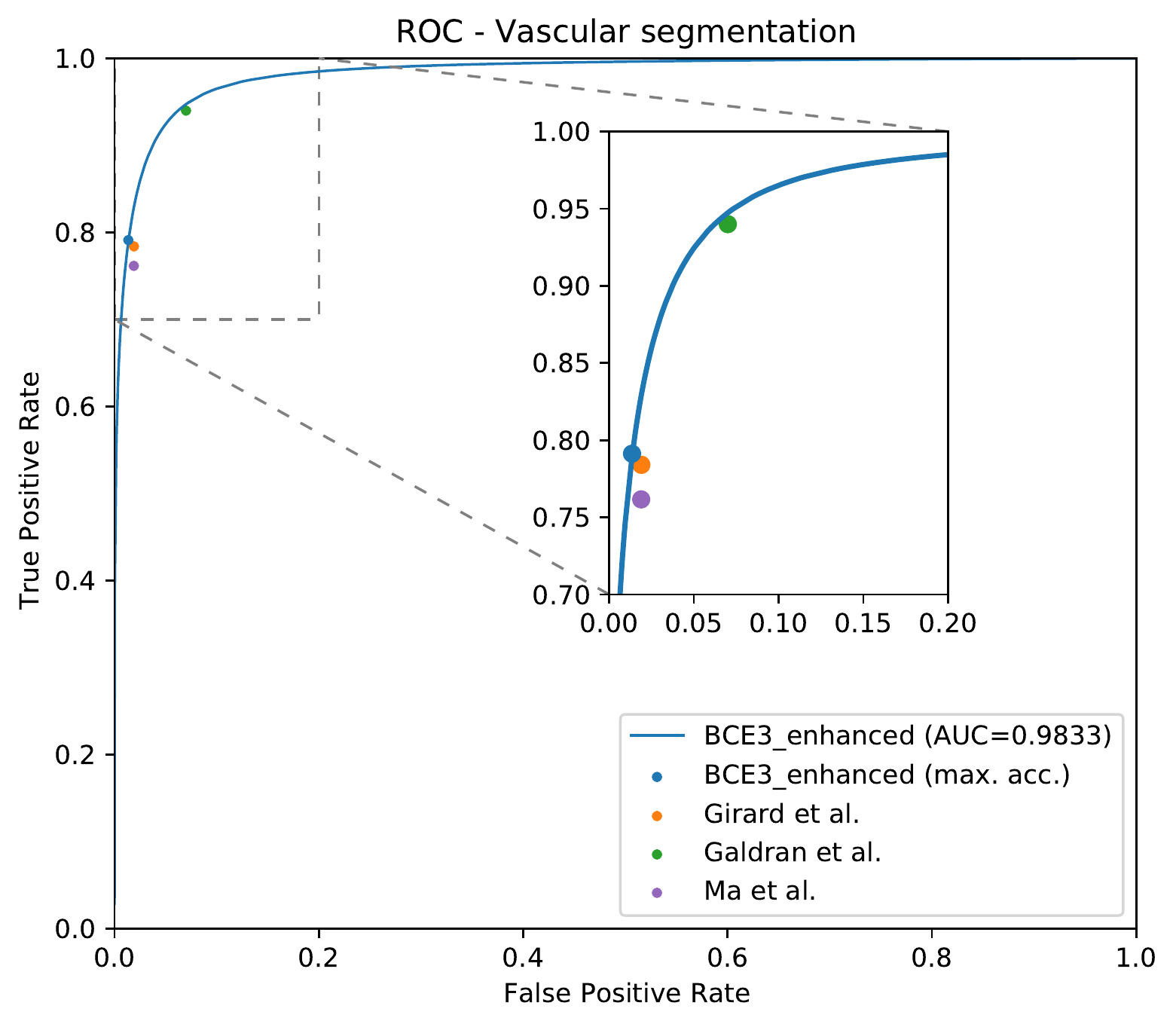}
        \caption{}
    \end{subfigure}
    \caption{ROC curves in the RITE-test set for our proposed system along with the point representations of the state-of-the-art approaches for the (a) artery/vein classification and (b) vascular segmentation tasks.}
    \label{fig:ROC_curves_SOTA}
\end{figure}

\begin{figure}
    \centering
    \includegraphics[width=0.6\linewidth]{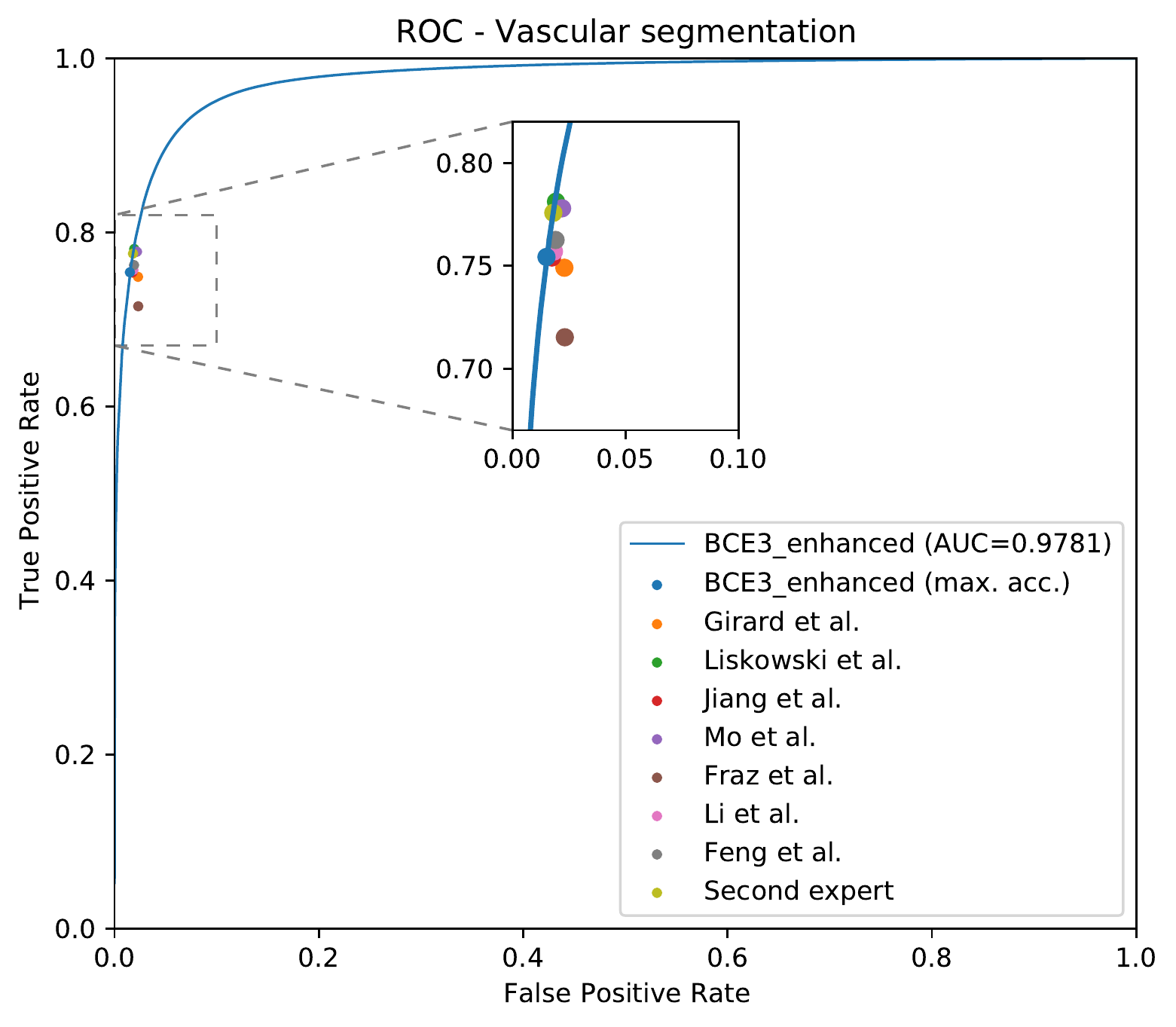}
    \caption{ROC curve in the DRIVE-test set for our proposed system along with the point representations of the state-of-the-art approaches for the vascular segmentation task.}
    \label{fig:roc_VO_mean_SOTA}
\end{figure}

Moreover, the methods in the state of the art report the results for a fixed operating point system, with a varying preference for false positives or false negatives among the works. As this complicates the direct comparison between the state-of-the-art results, we complement the results in Tables \ref{tab:classification_SOTA} and \ref{tab:vessels_segmentation_SOTA} with the ROC curves in Figure~\ref{fig:ROC_curves_SOTA}, for the artery/vein classification and the vascular segmentation tasks. In these graphs, we represent the ROC curves for our proposed system (BCE3 loss and enhanced retinographies as input) along with the point representations of the systems in Tables \ref{tab:classification_SOTA} and \ref{tab:vessels_segmentation_SOTA}.
Finally, in order to provide a context for the performance in the vasculature segmentation of the proposed system, we provide a comparison with reference works aiming at the vasculature segmentation in Table~\ref{tab:vessels_segmentation_SOTA}. Also, we complement the results shown in this table with the ROC curve depicted in Figure~\ref{fig:roc_VO_mean_SOTA}, as it was done for the RITE-test results. The reported results use the DRIVE-test set, which is composed of the same images as in the RITE-test set. However, in this case, the ground truth segmentations considered as gold standard are those provided by the first expert, and thus the obtained results are slightly different from the results reported for RITE.
\begin{table}
    \centering
    \caption{Comparison with the state of the art exclusively for the vessels segmentation task in DRIVE dataset. Here, the first expert is considered as gold standard. }
    \label{tab:vessels_segmentation_SOTA}

    \begin{tabular}{@{\extracolsep{4pt}}lllll}
    \hline
    \multicolumn{1}{l}{Method} & Sens. (\%) & Spec. (\%) & Acc. (\%) & AUC-ROC (\%) \\ \hline
    
    \multicolumn{1}{l}{Girard et al. \cite{Girard:Joint:AIM:2019}} & 74.9 & 97.7 & 94.8 & 96.4 \\ 
    \multicolumn{1}{l}{Liskowski et al. \cite{Liskowski:VS:TMI:2016}} & \textbf{78.11} & 98.07 & 95.35 & 97.90  \\ 
    \multicolumn{1}{l}{Jiang et al. \cite{Jiang:FullyConvVS:CMIG:2018}} & 75.40 $\pm$ 0.05 & 98.25 $\pm$ 0.01 & 96.24 $\pm$ 0.01 & \textbf{98.10} \\ 
    \multicolumn{1}{l}{Mo et al. \cite{Mo:IJCARS:2017}} & 77.79 & 97.80 & 95.21 & 97.82 \\ 
    \multicolumn{1}{l}{Fraz et al. \cite{Fraz:CMPB:2012}} & 71.52 & 97.68 & 94.30 $\pm$ 0.72 & - \\ 
    \multicolumn{1}{l}{Li et al. \cite{Li:TMI:2016}} & 75.69 & 98.16 & 95.27 & 97.38 \\ 
    \multicolumn{1}{l}{Feng et al. \cite{Feng:Neuro:2020}} & 76.25 & 98.09 & 95.28 & 96.78 \\ 
    \multicolumn{1}{l}{Hervella et al. \cite{Hervella:ASOC:2020}} & - & - & - & 97.82 \\ 
    \multicolumn{1}{l}{Second expert} & 77.57 & 98.19 & \textbf{96.37} & - \\ 
    \multicolumn{1}{l}{Proposed (enhanced)} & 75.42 $\pm$ 1.23 & \textbf{98.49 $\pm$ 0.13} & 95.45 $\pm$ 0.05 & 97.81 $\pm$ 0.02  \\ 
    \multicolumn{1}{l}{Proposed (original)} & 76.46 $\pm$ 0.52 & 98.36 $\pm$ 0.07 & 95.54 $\pm$ 0.02 & 97.81 $\pm$ 0.03  \\ \hline
    
    \end{tabular}%

    \end{table}

As it can be observed in the comparison results, the proposed approach achieves highly competitive results in both the artery/vein classification and the vascular segmentation tasks. In this regard, it is especially relevant that our approach achieves the best performance in the vascular segmentation task among the works aiming at solving both SSCAV tasks, and even a competitive performance with the state-of-the-art specialized methods. This is particularly relevant since, in a real A/V classification scenario, only the detected vessels would be classified into arteries and veins. Regarding the artery/vein classification of the ground truth vessels, however, the achieved results are slightly below the most recent works \cite{Ma:Multi-task:MICCAI:2019,Kang:A/V:CMPB:2020}.

In order to provide insight on the comparison results, it should be noticed that our method proposes a straightforward application of a plain U-Net architecture \cite{Ronneberger:U-Net:MICCAI:2015}, to full resolution images, by simply adjusting the used loss. 
Thus, the research herein described aims at providing methodological advances by simplifying the used methods, gaining insight on why they work, taking advantages of more appropriate ways of posing the end-to-end training of the target tasks, and avoiding the use of any bells and whistles.
In this regard, some of the prior works trained and applied the networks to a patch-wise \cite{Girard:Joint:AIM:2019,Ma:Multi-task:MICCAI:2019}, instead of using the full-size images as our case. We think that it is more convenient to take full advantage of the FCNN architectures to pose end-to-end training approaches that use full-size images as inputs, even if the patch-wise balancing and the augmentation approaches can take minor advantages on the results, depending on the used datasets.
Several methods, including the best performing ones, used \textit{ad hoc} weights to adjust the importance of each class on the loss \cite{Hemelings:A/V:CMIG:2019,Ma:Multi-task:MICCAI:2019,Kang:A/V:CMPB:2020}, or even heuristic weights to balance the importance of each vessel within each image \cite{Ma:Multi-task:MICCAI:2019,Kang:A/V:CMPB:2020}.
While we could adjust the class weights, as well as the importance of the difficult target structures, to improve the overall provided results of our approach, we think that such approach requires appropriate validation methodologies that are not usually practical giving the large training times of deep neural networks. There are also interesting ideas which can boost the performance of any approach, and could be potentially applied to our approach in future work, such as the use of pretrained networks \cite{Ma:Multi-task:MICCAI:2019,Kang:A/V:CMPB:2020} (ours are trained from scratch), enforcing the vascular tree classification coherence using an \textit{ad hoc} postprocessing approach \cite{Girard:Joint:AIM:2019,Kang:A/V:CMPB:2020}, or the use of deep supervision to enhance the backpropagation of gradients \cite{Ma:Multi-task:MICCAI:2019,Kang:A/V:CMPB:2020}. Notwithstanding we did not include any of these mechanisms, the achieved results are highly competitive in artery/vein classification, and certainly better in the vascular segmentation, which emphasizes the potential of the proposed approach.

The only additional mechanism from the state of the art that we included in our approach is the \textit{ad hoc} preprocessing. Initially, however, our intention was not to include it, since we consider that any neural network with an adequate capacity should be able to learn such simple preprocessing. Notwithstanding, its wide use in the literature and its significant impact in the classification results made us to finally incorporate one of these methods. In this sense, it is singular the contrast between the important improvement in classification and the almost null impact in segmentation. This observation could motivate future research.

Finally, it should be noticed that the proposed method presents an additional advantage over the rest of the state-of-the-art approaches: it allows to detect vessel crossings with a satisfactory performance. This aspect is distinctive of our work, since there is no other in the state of the art that allows to perform this task. Additionally, our approach gives raise to fully-connected arterial and venular trees. This factor facilitates the structural analysis of both types of vessels, and it is useful for the future application of methods for addressing the vascular coherence issue. Furthermore, it remarkable that these advantages are achieved with a simple change in the loss; a change that also simplifies the approach of the problem and significantly improves the performance on the vascular segmentation task.


\section{Conclusions}
\label{sec:conclusions}

In this work, we have proposed a novel approach for the simultaneous segmentation and classification of the retinal arteries and veins (SSCAV) using FCNNs. This approach decomposes this joint task into three segmentation problems: the segmentation of arteries, veins and the whole vascular tree. To train the networks following this approach, we have proposed a novel loss named BCE3, that combines the independent segmentation losses of the three classes of interest.

To evaluate the potential and advantages of this approach, we performed a comprehensive comparative study with the approach that currently represents the state of the art in the simultaneous segmentation and classification task. Contrary to ours, this approach uses Cross-Entropy as loss. Furthermore, we also compared the results achieved by our method with the current state-of-the-art results for both the SSCAV and exclusively the vascular segmentation task in the RITE and DRIVE datasets.

The results provided by the different experiments demonstrate that our proposal achieves a competitive performance in the classification of arteries and veins, as well as significantly improving the retinal vasculature segmentation. Complementarily, it allows to detect the complex vessel crossings with a satisfactory performance.

On the other hand, the comparison with the state of the art in the SSCAV and the vasculature segmentation tasks shows that our method achieves highly competitive results. In the SSCAV, it is remarkable that it surpasses the segmentation results of all the state-of-the-art approaches whereas preserving a high classification performance in comparison with the most recent methods. This aspect is valuable, as the current state of the art evaluate the vasculature classification on all the vessel pixels of the ground truth, but, in a real scenario, only the detected vessel pixels would be classified into arteries and veins. In this sense, the greater vessel detection effectiveness of our method represents a potential advantage. Furthermore, it is worth noting that these results are achieved with a direct adaptation and design of the used loss, and avoiding the use of any bells and whistles. In this regard, some of the mechanisms included in the most recent works (e.g. transfer learning, deep supervision or \textit{ad hoc} postprocessing) are compatible with our approach, and could probably improve its results.

In addition to the above, our method presents some inherent advantages. First, it is simple, since it consists of the straightforward application of a FCNN with a custom loss. Second, it provides continuous arterial and venular trees, also directly providing vasculature segmentation masks. Third, it allows to detect vessel crossings with an adequate effectiveness. These advantages are distinctive of our approach regarding to the current state of the art.

Notwithstanding, our approach also presents some potential aspects for further improvement. One of them, which also affects to the rest of the works that address the SSCAV, is the vessels classification incoherence. That is, the presence of manifestly misclassified segments of vessels in the output images. Although some works have proposed methods to mitigate these errors (see Section~\ref{subsec:SOTA}), they do not provide a fully learning-based approach capable of solving this problem for the most part of examples. With the current state-of-the-art methodologies, the classification between arteries and veins using FCNNs still relies on relatively low-level features. Thus, the problem of getting the networks to learn the high-level structure of the arteries and veins is still unsolved. Along with this issue, there is also room for improvement in the state of the art regarding another aspect: the image preprocessing. Although this technique demonstrated a positive impact in the A/V classification results, it does not improve the vascular segmentation. Also, it constitutes an \textit{ad hoc} stage that does not allow to train the networks in a truly end-to-end setting (from the original data to the target decisions). These two issues represent interesting fields for further research.


\section*{Declaration of competing interest}

The authors declare that they have no known competing financial interests or personal relationships that could have appeared to influence the work reported in this paper.


\section*{Acknowledgments}

This work is supported by Instituto de Salud Carlos III, Government of Spain, and the European Regional Development Fund (ERDF) of the European Union (EU) through the \mbox{DTS18/00136} research project;
Ministerio de Ciencia e Innovación, Government of Spain, through the \mbox{RTI2018-095894-B-I00} and \mbox{PID2019-108435RB-I00} research projects;
Xunta de Galicia and the European Social Fund (ESF) of the EU through the predoctoral grant contract ref. \mbox{ED481A-2017/328}; 
Consellería  de  Cultura,  Educación e Universidade, Xunta de Galicia, through Grupos de Referencia Competitiva, grant ref. ED431C 2020/24.
CITIC, Centro de Investigación de Galicia ref. \mbox{ED431G 2019/01}, receives financial support from Consellería de Educación, Universidade e Formación Profesional, Xunta de Galicia, through the ERDF (80\%) and Secretaría Xeral de Universidades (20\%).


\bibliography{references}

\begin{thebibliography}{10}
\expandafter\ifx\csname url\endcsname\relax
  \def\url#1{\texttt{#1}}\fi
\expandafter\ifx\csname urlprefix\endcsname\relax\def\urlprefix{URL }\fi
\expandafter\ifx\csname href\endcsname\relax
  \def\href#1#2{#2} \def\path#1{#1}\fi

\bibitem{Kanski:Ophthalmology:Elsevier:2011}
J.~J. Kanski, B.~Bowling, Clinical Ophthalmology: A Systematic Approach,
  seventh Edition, Elsevier Health Sciences, 2011.

\bibitem{Abramoff:Retinal:RBE:2010}
M.~D. {Abràmoff}, M.~K. {Garvin}, M.~{Sonka}, Retinal imaging and image
  analysis, IEEE Reviews in Biomedical Engineering 3 (2010) 169--208.
\newblock \href {https://doi.org/10.1109/RBME.2010.2084567}
  {\path{doi:10.1109/RBME.2010.2084567}}.

\bibitem{Rosenfeld:AMD:2006}
P.~J. Rosenfeld, D.~M. Brown, J.~S. Heier, D.~S. Boyer, P.~K. Kaiser, C.~Y.
  Chung, R.~Y. Kim, Ranibizumab for neovascular age-related macular
  degeneration, New England Journal of Medicine 355~(14) (2006) 1419--1431,
  pMID: 17021318.
\newblock \href {https://doi.org/10.1056/NEJMoa054481}
  {\path{doi:10.1056/NEJMoa054481}}.

\bibitem{daSilva:IJGM:2015}
A.~V.~B. da~Silva, S.~A. Gouvea, A.~P.~B. da~Silva, S.~Bortolon, A.~N.
  Rodrigues, G.~R. Abreu, F.~L. Herkenhoff, Changes in retinal microvascular
  diameter in patients with diabetes, International journal of general medicine
  8 (2015) 267--273.
\newblock \href {https://doi.org/10.2147/IJGM.S83749}
  {\path{doi:10.2147/IJGM.S83749}}.

\bibitem{Vandewalle:Glaucoma:AO:2014}
E.~Vandewalle, L.~Abegão~Pinto, O.~B. Olafsdottir, E.~De~Clerck, P.~Stalmans,
  J.~Van~Calster, T.~Zeyen, E.~Stefánsson, I.~Stalmans, Oximetry in glaucoma:
  correlation of metabolic change with structural and functional damage, Acta
  Ophthalmologica 92~(2) (2014) 105--110.
\newblock \href {https://doi.org/10.1111/aos.12011}
  {\path{doi:10.1111/aos.12011}}.

\bibitem{Ramm:Vessels:GACEO:2014}
L.~Ramm, S.~Jentsch, S.~Peters, R.~Augsten, M.~Hammer, Investigation of blood
  flow regulation and oxygen saturation of the retinal vessels in primary
  open-angle glaucoma, Graefe's Archive for Clinical and Experimental
  Ophthalmology 252~(11) (2014) 1803--1810.
\newblock \href {https://doi.org/10.1007/s00417-014-2766-4}
  {\path{doi:10.1007/s00417-014-2766-4}}.

\bibitem{Wang:Diameter:CEO:2007}
S.~Wang, L.~Xu, Y.~Wang, Y.~Wang, J.~B. Jonas, {{R}etinal vessel diameter in
  normal and glaucomatous eyes: the {B}eijing eye study}, Clin. Experiment.
  Ophthalmol. 35~(9) (2007) 800--807.
\newblock \href {https://doi.org/10.1111/j.1442-9071.2007.01627.x}
  {\path{doi:10.1111/j.1442-9071.2007.01627.x}}.

\bibitem{Mitchell:Diameters:O:2005}
P.~Mitchell, H.~Leung, J.~Wang, E.~Rochtchina, A.~Lee, T.~Wong, R.~Klein,
  Retinal vessel diameter and open-angle glaucoma: The blue mountains eye
  study, Ophthalmology 112~(2) (2005) 245--250, cited By 162.
\newblock \href {https://doi.org/10.1016/j.ophtha.2004.08.015}
  {\path{doi:10.1016/j.ophtha.2004.08.015}}.

\bibitem{Mathers:Diseases:2006}
C.~Mathers, D.~Loncar, Projections of global mortality and burden of disease
  from 2002 to 2030, PLoS Medicine 3~(11) (2006) 2011--2030, cited By 5959.
\newblock \href {https://doi.org/10.1371/journal.pmed.0030442}
  {\path{doi:10.1371/journal.pmed.0030442}}.

\bibitem{Ross:Atherosclerosis:1999}
R.~Ross, Atherosclerosis - an inflammatory disease, New England Journal of
  Medicine 340~(2) (1999) 115--126, cited By 17782.
\newblock \href {https://doi.org/10.1056/NEJM199901143400207}
  {\path{doi:10.1056/NEJM199901143400207}}.

\bibitem{Prokofyeva:Epidemiology:2012}
E.~Prokofyeva, E.~Zrenner, Epidemiology of major eye diseases leading to
  blindness in europe: A literature review, Ophthalmic Research 47~(4) (2012)
  171--188.
\newblock \href {https://doi.org/10.1159/000329603}
  {\path{doi:10.1159/000329603}}.

\bibitem{Staal:DRIVE:2004}
J.~{Staal}, M.~D. {Abràmoff}, M.~{Niemeijer}, M.~A. {Viergever}, B.~{van
  Ginneken}, Ridge-based vessel segmentation in color images of the retina,
  IEEE Transactions on Medical Imaging 23~(4) (2004) 501--509.
\newblock \href {https://doi.org/10.1109/TMI.2004.825627}
  {\path{doi:10.1109/TMI.2004.825627}}.

\bibitem{Patton:Retinal:JAnat:2005}
N.~Patton, T.~Aslam, T.~Macgillivray, A.~Pattie, I.~J. Deary, B.~Dhillon,
  Retinal vascular image analysis as a potential screening tool for
  cerebrovascular disease: a rationale based on homology between cerebral and
  retinal microvasculatures, Journal of anatomy 206~(4) (2005) 319--348.
\newblock \href {https://doi.org/10.1111/j.1469-7580.2005.00395.x}
  {\path{doi:10.1111/j.1469-7580.2005.00395.x}}.

\bibitem{Fraz:Segmentation:CMPB:2012}
M.~Fraz, P.~Remagnino, A.~Hoppe, B.~Uyyanonvara, A.~Rudnicka, C.~Owen,
  S.~Barman, Blood vessel segmentation methodologies in retinal images – a
  survey, Computer Methods and Programs in Biomedicine 108~(1) (2012) 407 --
  433.
\newblock \href {https://doi.org/10.1016/j.cmpb.2012.03.009}
  {\path{doi:10.1016/j.cmpb.2012.03.009}}.

\bibitem{Kawasaki:Hypertension:JH:2009}
R.~Kawasaki, N.~Cheung, J.~J. Wang, R.~Klein, B.~E. Klein, M.~F. Cotch, A.~R.
  Sharrett, S.~Shea, F.~A. Islam, T.~Y. Wong, Retinal vessel diameters and risk
  of hypertension: the multiethnic study of atherosclerosis, Journal of
  Hypertension 27~(12) (2009) 2386--2393.
\newblock \href {https://doi.org/10.1097/HJH.0b013e3283310f7e}
  {\path{doi:10.1097/HJH.0b013e3283310f7e}}.

\bibitem{Ikram:AVR:IOVS:2004}
M.~K. Ikram, F.~J. de~Jong, J.~R. Vingerling, J.~C.~M. Witteman, A.~Hofman,
  M.~M.~B. Breteler, P.~T. V.~M. de~Jong, {Are Retinal Arteriolar or Venular
  Diameters Associated with Markers for Cardiovascular Disorders? The Rotterdam
  Study}, Investigative Ophthalmology \& Visual Science 45~(7) (2004)
  2129--2134.
\newblock \href {https://doi.org/10.1167/iovs.03-1390}
  {\path{doi:10.1167/iovs.03-1390}}.

\bibitem{Sun:AVR:SO:2009}
C.~Sun, J.~J. Wang, D.~A. Mackey, T.~Y. Wong, Retinal vascular caliber:
  Systemic, environmental, and genetic associations, Survey of Ophthalmology
  54~(1) (2009) 74 -- 95.
\newblock \href {https://doi.org/10.1016/j.survophthal.2008.10.003}
  {\path{doi:10.1016/j.survophthal.2008.10.003}}.

\bibitem{Hatanaka:Narrowing:EMB:2005}
Y.~{Hatanaka}, T.~{Nakagawa}, A.~{Aoyama}, X.~{Zhou}, T.~{Hara}, H.~{Fujita},
  M.~{Kakogawa}, Y.~{Hayashi}, Y.~{Mizukusa}, A.~{Fujita}, Automated detection
  algorithm for arteriolar narrowing on fundus images, in: 2005 IEEE
  Engineering in Medicine and Biology 27th Annual Conference, 2005, pp.
  286--289.
\newblock \href {https://doi.org/10.1109/IEMBS.2005.1616400}
  {\path{doi:10.1109/IEMBS.2005.1616400}}.

\bibitem{Jiang:Adaptive:TPAMI:2003}
X.~{Jiang}, D.~{Mojon}, Adaptive local thresholding by verification-based
  multithreshold probing with application to vessel detection in retinal
  images, IEEE Transactions on Pattern Analysis and Machine Intelligence 25~(1)
  (2003) 131--137.
\newblock \href {https://doi.org/10.1109/TPAMI.2003.1159954}
  {\path{doi:10.1109/TPAMI.2003.1159954}}.

\bibitem{Nain:Shape:MICCAI:2004}
D.~Nain, A.~Yezzi, G.~Turk, Vessel segmentation using a shape driven flow, in:
  Medical Image Computing and Computer-Assisted Intervention -- MICCAI, Vol.
  3216 of LNCS, 2004, pp. 51--59.
\newblock \href {https://doi.org/10.1007/978-3-540-30135-6_7}
  {\path{doi:10.1007/978-3-540-30135-6_7}}.

\bibitem{Tolias:Fuzzy:TMI:1998}
Y.~A. {Tolias}, S.~M. {Panas}, A fuzzy vessel tracking algorithm for retinal
  images based on fuzzy clustering, IEEE Transactions on Medical Imaging 17~(2)
  (1998) 263--273.
\newblock \href {https://doi.org/10.1109/42.700738}
  {\path{doi:10.1109/42.700738}}.

\bibitem{Sinthanayothin:Localisation:BJO:1999}
C.~Sinthanayothin, J.~F. Boyce, H.~L. Cook, T.~H. Williamson, Automated
  localisation of the optic disc, fovea, and retinal blood vessels from digital
  colour fundus images, British Journal of Ophthalmology 83~(8) (1999)
  902--910.
\newblock \href {https://doi.org/10.1136/bjo.83.8.902}
  {\path{doi:10.1136/bjo.83.8.902}}.

\bibitem{Marin:SupervisedVS:TMI:2011}
D.~{Marín}, A.~{Aquino}, M.~E. {Gegundez-Arias}, J.~M. {Bravo}, A new
  supervised method for blood vessel segmentation in retinal images by using
  gray-level and moment invariants-based features, IEEE Transactions on Medical
  Imaging 30~(1) (2011) 146--158.
\newblock \href {https://doi.org/10.1109/TMI.2010.2064333}
  {\path{doi:10.1109/TMI.2010.2064333}}.

\bibitem{Liskowski:VS:TMI:2016}
P.~{Liskowski}, K.~{Krawiec}, Segmenting retinal blood vessels with deep neural
  networks, IEEE Transactions on Medical Imaging 35~(11) (2016) 2369--2380.
\newblock \href {https://doi.org/10.1109/TMI.2016.2546227}
  {\path{doi:10.1109/TMI.2016.2546227}}.

\bibitem{Fu:DeepVessel:MICCAI:2016}
H.~Fu, Y.~Xu, S.~Lin, D.~W. Kee~Wong, J.~Liu, Deep{V}essel: Retinal vessel
  segmentation via deep learning and conditional random field, in:
  S.~Ourselin, L.~Joskowicz, M.~R. Sabuncu, G.~Unal, W.~Wells (Eds.), Medical
  Image Computing and Computer-Assisted Intervention -- MICCAI 2016, Springer
  International Publishing, Cham, 2016, pp. 132--139.
\newblock \href {https://doi.org/10.1007/978-3-319-46723-8_16}
  {\path{doi:10.1007/978-3-319-46723-8_16}}.

\bibitem{Fu:VSFully:ISBI:2016}
H.~{Fu}, Y.~{Xu}, D.~W.~K. {Wong}, J.~{Liu}, Retinal vessel segmentation via
  deep learning network and fully-connected conditional random fields, in: 2016
  IEEE 13th International Symposium on Biomedical Imaging (ISBI), 2016, pp.
  698--701.
\newblock \href {https://doi.org/10.1109/ISBI.2016.7493362}
  {\path{doi:10.1109/ISBI.2016.7493362}}.

\bibitem{Dasgupta:VesselSegmentationFCNN:ISBI:2017}
A.~{Dasgupta}, S.~{Singh}, A fully convolutional neural network based
  structured prediction approach towards the retinal vessel segmentation, in:
  2017 IEEE 14th International Symposium on Biomedical Imaging (ISBI), 2017,
  pp. 248--251.
\newblock \href {https://doi.org/10.1109/ISBI.2017.7950512}
  {\path{doi:10.1109/ISBI.2017.7950512}}.

\bibitem{Jiang:FullyConvVS:CMIG:2018}
Z.~Jiang, H.~Zhang, Y.~Wang, S.-B. Ko, Retinal blood vessel segmentation using
  fully convolutional network with transfer learning, Computerized Medical
  Imaging and Graphics 68 (2018) 1 -- 15.
\newblock \href {https://doi.org/10.1016/j.compmedimag.2018.04.005}
  {\path{doi:10.1016/j.compmedimag.2018.04.005}}.

\bibitem{Oliveira:FullyConvVS:ESA:2018}
A.~Oliveira, S.~Pereira, C.~A. Silva, Retinal vessel segmentation based on
  fully convolutional neural networks, Expert Systems with Applications 112
  (2018) 229 -- 242.
\newblock \href {https://doi.org/10.1016/j.eswa.2018.06.034}
  {\path{doi:10.1016/j.eswa.2018.06.034}}.

\bibitem{Jin:DUNet:KBS:2019}
Q.~Jin, Z.~Meng, T.~D. Pham, Q.~Chen, L.~Wei, R.~Su, {DUNet}: A deformable
  network for retinal vessel segmentation, Knowledge-Based Systems 178 (2019)
  149 -- 162.
\newblock \href {https://doi.org/10.1016/j.knosys.2019.04.025}
  {\path{doi:10.1016/j.knosys.2019.04.025}}.

\bibitem{Relan:Retinal:EMBC:2013}
D.~{Relan}, T.~{MacGillivray}, L.~{Ballerini}, E.~{Trucco}, Retinal vessel
  classification: Sorting arteries and veins, in: 2013 35th Annual
  International Conference of the IEEE Engineering in Medicine and Biology
  Society (EMBC), 2013, pp. 7396--7399.
\newblock \href {https://doi.org/10.1109/EMBC.2013.6611267}
  {\path{doi:10.1109/EMBC.2013.6611267}}.

\bibitem{Relan:EMBS:2014}
D.~{Relan}, T.~{MacGillivray}, L.~{Ballerini}, E.~{Trucco}, Automatic retinal
  vessel classification using a least square-support vector machine in vampire,
  in: 2014 36th Annual International Conference of the IEEE Engineering in
  Medicine and Biology Society, 2014, pp. 142--145.
\newblock \href {https://doi.org/10.1109/EMBC.2014.6943549}
  {\path{doi:10.1109/EMBC.2014.6943549}}.

\bibitem{Dashtbozorg:TIP:2014}
B.~{Dashtbozorg}, A.~M. {Mendonça}, A.~{Campilho}, An automatic graph-based
  approach for artery/vein classification in retinal images, IEEE Transactions
  on Image Processing 23~(3) (2014) 1073--1083.
\newblock \href {https://doi.org/10.1109/TIP.2013.2263809}
  {\path{doi:10.1109/TIP.2013.2263809}}.

\bibitem{Estrada:TMI:2015}
R.~{Estrada}, M.~J. {Allingham}, P.~S. {Mettu}, S.~W. {Cousins}, C.~{Tomasi},
  S.~{Farsiu}, Retinal artery-vein classification via topology estimation, IEEE
  Transactions on Medical Imaging 34~(12) (2015) 2518--2534.
\newblock \href {https://doi.org/10.1109/TMI.2015.2443117}
  {\path{doi:10.1109/TMI.2015.2443117}}.

\bibitem{Welikala:CBM:2017}
R.~Welikala, P.~Foster, P.~Whincup, A.~Rudnicka, C.~Owen, D.~Strachan,
  S.~Barman, Automated arteriole and venule classification using deep learning
  for retinal images from the uk biobank cohort, Computers in Biology and
  Medicine 90 (2017) 23 -- 32.
\newblock \href {https://doi.org/10.1016/j.compbiomed.2017.09.005}
  {\path{doi:10.1016/j.compbiomed.2017.09.005}}.

\bibitem{Xu:A/V:BOE:2018}
X.~Xu, R.~Wang, P.~Lv, B.~Gao, C.~Li, Z.~Tian, T.~Tan, F.~Xu, Simultaneous
  arteriole and venule segmentation with domain-specific loss function on a new
  public database, Biomedical Optics Express 9~(7) (2018) 3153--3166.
\newblock \href {https://doi.org/10.1364/BOE.9.003153}
  {\path{doi:10.1364/BOE.9.003153}}.

\bibitem{Girard:Joint:AIM:2019}
F.~Girard, C.~Kavalec, F.~Cheriet, Joint segmentation and classification of
  retinal arteries/veins from fundus images, Artificial Intelligence in
  Medicine 94 (2019) 96 -- 109.
\newblock \href {https://doi.org/10.1016/j.artmed.2019.02.004}
  {\path{doi:10.1016/j.artmed.2019.02.004}}.

\bibitem{Hemelings:A/V:CMIG:2019}
R.~Hemelings, B.~Elen, I.~Stalmans, K.~{Van Keer}, P.~{De Boever}, M.~B.
  Blaschko, Artery–vein segmentation in fundus images using a fully
  convolutional network, Computerized Medical Imaging and Graphics 76 (2019)
  101636.
\newblock \href {https://doi.org/10.1016/j.compmedimag.2019.05.004}
  {\path{doi:10.1016/j.compmedimag.2019.05.004}}.

\bibitem{Galdran:Uncertainty:ISBI:2019}
A.~{Galdran}, M.~{Meyer}, P.~{Costa}, {MendonÇa}, A.~{Campilho},
  Uncertainty-aware artery/vein classification on retinal images, in: 2019 IEEE
  16th International Symposium on Biomedical Imaging (ISBI 2019), 2019, pp.
  556--560.
\newblock \href {https://doi.org/10.1109/ISBI.2019.8759380}
  {\path{doi:10.1109/ISBI.2019.8759380}}.

\bibitem{Ma:Multi-task:MICCAI:2019}
W.~Ma, S.~Yu, K.~Ma, J.~Wang, X.~Ding, Y.~Zheng, Multi-task neural networks
  with spatial activation for retinal vessel segmentation and artery/vein
  classification, in: D.~Shen, T.~Liu, T.~M. Peters, L.~H. Staib, C.~Essert,
  S.~Zhou, P.-T. Yap, A.~Khan (Eds.), Medical Image Computing and Computer
  Assisted Intervention -- MICCAI 2019, Springer International Publishing,
  Cham, 2019, pp. 769--778.
\newblock \href {https://doi.org/10.1007/978-3-030-32239-7_85}
  {\path{doi:10.1007/978-3-030-32239-7_85}}.

\bibitem{Kang:A/V:CMPB:2020}
H.~Kang, Y.~Gao, S.~Guo, X.~Xu, T.~Li, K.~Wang, {AVN}et: A retinal artery/vein
  classification network with category-attention weighted fusion, Computer
  Methods and Programs in Biomedicine 195 (2020) 105629.
\newblock \href {https://doi.org/10.1016/j.cmpb.2020.105629}
  {\path{doi:10.1016/j.cmpb.2020.105629}}.

\bibitem{Hu:RITE:2013}
Q.~Hu, M.~D. Abr{\`a}moff, M.~K. Garvin, Automated separation of binary
  overlapping trees in low-contrast color retinal images, in: K.~Mori,
  I.~Sakuma, Y.~Sato, C.~Barillot, N.~Navab (Eds.), Medical Image Computing and
  Computer-Assisted Intervention -- MICCAI 2013, Springer Berlin Heidelberg,
  Berlin, Heidelberg, 2013, pp. 436--443.
\newblock \href {https://doi.org/10.1007/978-3-642-40763-5_54}
  {\path{doi:10.1007/978-3-642-40763-5_54}}.

\bibitem{Qureshi:ISCBMS:2013}
T.~A. {Qureshi}, M.~{Habib}, A.~{Hunter}, B.~{Al-Diri}, A manually-labeled,
  artery/vein classified benchmark for the drive dataset, in: Proceedings of
  the 26th IEEE International Symposium on Computer-Based Medical Systems,
  2013, pp. 485--488.
\newblock \href {https://doi.org/10.1109/CBMS.2013.6627847}
  {\path{doi:10.1109/CBMS.2013.6627847}}.

\bibitem{Orlando:MICCAI:2018}
J.~I. Orlando, J.~Barbosa~Breda, K.~van Keer, M.~B. Blaschko, P.~J. Blanco,
  C.~A. Bulant, Towards a glaucoma risk index based on simulated hemodynamics
  from fundus images, in: A.~F. Frangi, J.~A. Schnabel, C.~Davatzikos,
  C.~Alberola-L{\'o}pez, G.~Fichtinger (Eds.), Medical Image Computing and
  Computer Assisted Intervention -- MICCAI 2018, Springer International
  Publishing, Cham, 2018, pp. 65--73.
\newblock \href {https://doi.org/10.1007/978-3-030-00934-2_8}
  {\path{doi:10.1007/978-3-030-00934-2_8}}.

\bibitem{Rothaus:IVC:2009}
K.~Rothaus, X.~Jiang, P.~Rhiem, Separation of the retinal vascular graph in
  arteries and veins based upon structural knowledge, Image and Vision
  Computing 27~(7) (2009) 864 -- 875, 7th IAPR-TC15 Workshop on Graph-based
  Representations (GbR 2007).
\newblock \href {https://doi.org/10.1016/j.imavis.2008.02.013}
  {\path{doi:10.1016/j.imavis.2008.02.013}}.

\bibitem{LeCun:Gradient:IEEE:1998}
Y.~{LeCun}, L.~Bottou, Y.~Bengio, P.~Haffner, Gradient-based learning applied
  to document recognition, in: Proceedings of the IEEE, 1998, pp. 2278--2324.
\newblock \href {https://doi.org/10.1109/5.726791}
  {\path{doi:10.1109/5.726791}}.

\bibitem{Krizhevsky:AlexNet:NIPS:2012}
A.~Krizhevsky, I.~Sutskever, G.~E. Hinton, Image{N}et classification with deep
  convolutional neural networks, in: Proceedings of the 25th International
  Conference on Neural Information Processing Systems - Volume 1, NIPS'12,
  Curran Associates Inc., Red Hook, NY, USA, 2012, p. 1097–1105.
\newblock \href {https://doi.org/10.1145/3065386} {\path{doi:10.1145/3065386}}.

\bibitem{Long:FullyConv:CVPR:2015}
J.~{Long}, E.~{Shelhamer}, T.~{Darrell}, Fully convolutional networks for
  semantic segmentation, in: 2015 IEEE Conference on Computer Vision and
  Pattern Recognition (CVPR), 2015, pp. 3431--3440.
\newblock \href {https://doi.org/10.1109/CVPR.2015.7298965}
  {\path{doi:10.1109/CVPR.2015.7298965}}.

\bibitem{Ronneberger:U-Net:MICCAI:2015}
O.~Ronneberger, P.~Fischer, T.~Brox, U-net: Convolutional networks for
  biomedical image segmentation, in: N.~Navab, J.~Hornegger, W.~M. Wells, A.~F.
  Frangi (Eds.), Medical Image Computing and Computer-Assisted Intervention --
  MICCAI 2015, Springer International Publishing, Cham, 2015, pp. 234--241.
\newblock \href {https://doi.org/10.1007/978-3-319-24574-4_28}
  {\path{doi:10.1007/978-3-319-24574-4_28}}.

\bibitem{Zamperini:Features:CBMS:2012}
A.~{Zamperini}, A.~{Giachetti}, E.~{Trucco}, K.~S. {Chin}, Effective features
  for artery-vein classification in digital fundus images, in: 2012 25th IEEE
  International Symposium on Computer-Based Medical Systems (CBMS), 2012, pp.
  1--6.
\newblock \href {https://doi.org/10.1109/CBMS.2012.6266336}
  {\path{doi:10.1109/CBMS.2012.6266336}}.

\bibitem{Saez:CMPB:2012}
M.~Saez, S.~González-Vázquez, M.~González-Penedo, M.~A. Barceló,
  M.~Pena-Seijo, G.~{Coll de Tuero}, A.~Pose-Reino, Development of an automated
  system to classify retinal vessels into arteries and veins, Computer Methods
  and Programs in Biomedicine 108~(1) (2012) 367 -- 376.
\newblock \href {https://doi.org/10.1016/j.cmpb.2012.02.008}
  {\path{doi:10.1016/j.cmpb.2012.02.008}}.

\bibitem{Hao:Neurocomputing:2020}
S.~Hao, Y.~Zhou, Y.~Guo, A brief survey on semantic segmentation with deep
  learning, Neurocomputing 406 (2020) 302 -- 321.
\newblock \href {https://doi.org/10.1016/j.neucom.2019.11.118}
  {\path{doi:10.1016/j.neucom.2019.11.118}}.

\bibitem{Szegedy:CVPR:2015}
C.~{Szegedy}, {Wei Liu}, {Yangqing Jia}, P.~{Sermanet}, S.~{Reed},
  D.~{Anguelov}, D.~{Erhan}, V.~{Vanhoucke}, A.~{Rabinovich}, Going deeper with
  convolutions, in: 2015 IEEE Conference on Computer Vision and Pattern
  Recognition (CVPR), 2015, pp. 1--9.
\newblock \href {https://doi.org/10.1109/CVPR.2015.7298594}
  {\path{doi:10.1109/CVPR.2015.7298594}}.

\bibitem{He:CVPR:2016}
K.~{He}, X.~{Zhang}, S.~{Ren}, J.~{Sun}, Deep residual learning for image
  recognition, in: 2016 IEEE Conference on Computer Vision and Pattern
  Recognition (CVPR), 2016, pp. 770--778.
\newblock \href {https://doi.org/10.1109/CVPR.2016.90}
  {\path{doi:10.1109/CVPR.2016.90}}.

\bibitem{Zhu:NIPS:2017}
J.-Y. Zhu, R.~Zhang, D.~Pathak, T.~Darrell, A.~A. Efros, O.~Wang, E.~Shechtman,
  Toward multimodal image-to-image translation, in: I.~Guyon, U.~V. Luxburg,
  S.~Bengio, H.~Wallach, R.~Fergus, S.~Vishwanathan, R.~Garnett (Eds.),
  Advances in Neural Information Processing Systems 30, Curran Associates,
  Inc., 2017, pp. 465--476.

\bibitem{Isola:CVPR:2017}
P.~{Isola}, J.~{Zhu}, T.~{Zhou}, A.~A. {Efros}, Image-to-image translation with
  conditional adversarial networks, in: 2017 IEEE Conference on Computer Vision
  and Pattern Recognition (CVPR), 2017, pp. 5967--5976.
\newblock \href {https://doi.org/10.1109/CVPR.2017.632}
  {\path{doi:10.1109/CVPR.2017.632}}.

\bibitem{Falk:Cell:NM:2019}
T.~Falk, D.~Mai, R.~Bensch, {\"O}.~{\c{C}}i{\c{c}}ek, A.~Abdulkadir,
  Y.~Marrakchi, A.~B{\"o}hm, J.~Deubner, Z.~J{\"a}ckel, K.~Seiwald,
  A.~Dovzhenko, O.~Tietz, C.~Dal~Bosco, S.~Walsh, D.~Saltukoglu, T.~L. Tay,
  M.~Prinz, K.~Palme, M.~Simons, I.~Diester, T.~Brox, O.~Ronneberger, U-net:
  deep learning for cell counting, detection, and morphometry, Nature Methods
  16~(1) (2019) 67--70.
\newblock \href {https://doi.org/10.1038/s41592-018-0261-2}
  {\path{doi:10.1038/s41592-018-0261-2}}.

\bibitem{Fu:Optic:TMI:2018}
H.~{Fu}, J.~{Cheng}, Y.~{Xu}, D.~W.~K. {Wong}, J.~{Liu}, X.~{Cao}, Joint optic
  disc and cup segmentation based on multi-label deep network and polar
  transformation, IEEE Transactions on Medical Imaging 37~(7) (2018)
  1597--1605.
\newblock \href {https://doi.org/10.1109/TMI.2018.2791488}
  {\path{doi:10.1109/TMI.2018.2791488}}.

\bibitem{Litjens:MIA:2017}
G.~Litjens, T.~Kooi, B.~E. Bejnordi, A.~A.~A. Setio, F.~Ciompi, M.~Ghafoorian,
  J.~A. {van der Laak}, B.~{van Ginneken}, C.~I. Sánchez, A survey on deep
  learning in medical image analysis, Medical Image Analysis 42 (2017) 60 --
  88.
\newblock \href {https://doi.org/10.1016/j.media.2017.07.005}
  {\path{doi:10.1016/j.media.2017.07.005}}.

\bibitem{Morano:Segmentation:ECAI:2020}
J.~Morano, {\'A}.~S. Hervella, N.~Barreira, J.~Novo, J.~Rouco, Multimodal
  transfer learning-based approaches for retinal vascular segmentation, in:
  European Conference on Artificial Intelligence (ECAI), Vol. 325 of Frontiers
  in Artificial Intelligence and Applications, 2020, pp. 1866--1873.
\newblock \href {https://doi.org/10.3233/FAIA200303}
  {\path{doi:10.3233/FAIA200303}}.

\bibitem{Hervella:ASOC:2020}
Álvaro S.~Hervella, J.~Rouco, J.~Novo, M.~Ortega, Learning the retinal anatomy
  from scarce annotated data using self-supervised multimodal reconstruction,
  Applied Soft Computing 91 (2020) 106210.
\newblock \href {https://doi.org/10.1016/j.asoc.2020.106210}
  {\path{doi:10.1016/j.asoc.2020.106210}}.

\bibitem{Hervella:ESWA:2020}
Álvaro S.~Hervella, J.~Rouco, J.~Novo, M.~Ortega, Self-supervised multimodal
  reconstruction of retinal images over paired datasets, Expert Systems with
  Applications 161 (2020) 113674.
\newblock \href {https://doi.org/10.1016/j.eswa.2020.113674}
  {\path{doi:10.1016/j.eswa.2020.113674}}.

\bibitem{Kingma:Adam:2015}
D.~P. Kingma, J.~Ba, Adam: {A} method for stochastic optimization, in: 3rd
  International Conference on Learning Representations, {ICLR}, San Diego, CA,
  USA, May 7-9, 2015, Conference Track Proceedings, 2015.

\bibitem{Hervella:CMPB:2020}
Álvaro S.~Hervella, J.~Rouco, J.~Novo, M.~G. Penedo, M.~Ortega, Deep
  multi-instance heatmap regression for the detection of retinal vessel
  crossings and bifurcations in eye fundus images, Computer Methods and
  Programs in Biomedicine 186 (2020) 105201.
\newblock \href {https://doi.org/10.1016/j.cmpb.2019.105201}
  {\path{doi:10.1016/j.cmpb.2019.105201}}.

\bibitem{He:Initialization:2015}
K.~He, X.~Zhang, S.~Ren, J.~Sun, Delving deep into rectifiers: Surpassing
  human-level performance on imagenet classification, in: Proceedings of the
  2015 IEEE International Conference on Computer Vision (ICCV), ICCV,
  Washington, DC, USA, 2015, pp. 1026--1034.
\newblock \href {https://doi.org/10.1109/ICCV.2015.123}
  {\path{doi:10.1109/ICCV.2015.123}}.

\bibitem{Davis:ICML:2006}
J.~Davis, M.~Goadrich, The relationship between precision-recall and roc
  curves, in: Proceedings of the 23rd International Conference on Machine
  Learning, ACM, ICML '06, Association for Computing Machinery, New York, NY,
  USA, 2006, p. 233–240.
\newblock \href {https://doi.org/10.1145/1143844.1143874}
  {\path{doi:10.1145/1143844.1143874}}.

\bibitem{Mo:IJCARS:2017}
J.~Mo, L.~Zhang, Multi-level deep supervised networks for retinal vessel
  segmentation, International Journal of Computer Assisted Radiology and
  Surgery 12~(12) (2017) 2181--2193.
\newblock \href {https://doi.org/10.1007/s11548-017-1619-0}
  {\path{doi:10.1007/s11548-017-1619-0}}.

\bibitem{Fraz:CMPB:2012}
M.~Fraz, S.~Barman, P.~Remagnino, A.~Hoppe, A.~Basit, B.~Uyyanonvara,
  A.~Rudnicka, C.~Owen, An approach to localize the retinal blood vessels using
  bit planes and centerline detection, Computer Methods and Programs in
  Biomedicine 108~(2) (2012) 600 -- 616.
\newblock \href {https://doi.org/10.1016/j.cmpb.2011.08.009}
  {\path{doi:10.1016/j.cmpb.2011.08.009}}.

\bibitem{Li:TMI:2016}
Q.~{Li}, B.~{Feng}, L.~{Xie}, P.~{Liang}, H.~{Zhang}, T.~{Wang}, A
  cross-modality learning approach for vessel segmentation in retinal images,
  IEEE Transactions on Medical Imaging 35~(1) (2016) 109--118.
\newblock \href {https://doi.org/10.1109/TMI.2015.2457891}
  {\path{doi:10.1109/TMI.2015.2457891}}.

\bibitem{Feng:Neuro:2020}
S.~Feng, Z.~Zhuo, D.~Pan, Q.~Tian, Ccnet: A cross-connected convolutional
  network for segmenting retinal vessels using multi-scale features,
  Neurocomputing 392 (2020) 268 -- 276.
\newblock \href {https://doi.org/10.1016/j.neucom.2018.10.098}
  {\path{doi:10.1016/j.neucom.2018.10.098}}.

\end{thebibliography}

\end{document}